\documentclass[10pt]{article}
\usepackage{tgpagella}
\usepackage[dvipsnames]{xcolor}
\usepackage{authblk}
\usepackage{amsmath, mathtools}
\usepackage{amsfonts}
\usepackage{braket}
\usepackage{siunitx}
\usepackage{graphicx}
\usepackage{multicol}
\usepackage{caption}
\usepackage{amssymb}
\usepackage{tabulary}
\usepackage{booktabs}
\usepackage{array}
\usepackage[parfill]{parskip}
\usepackage{tikz}
\usepackage[margin=0.5in]{geometry}
\usepackage{textcomp}
\usepackage{subfig}
\usepackage{tensind}
\tensorformat{lrb}
\tensordelimiter{?}
\usepackage{float}
\usepackage{pgfplots}
\usepackage{threeparttable}
\usepackage{notoccite}
\usepackage[backend=biber, sorting=none, style=chem-acs, doi=false]{biblatex}
\renewbibmacro{in:}{}
\bibliography{Bibliography}

\pgfplotsset{every axis/.append style={very thick,  tick style={ultra thick}}, compat=1.9, width=8.7cm}
\usetikzlibrary{pgfplots.groupplots}

\usetikzlibrary{matrix,positioning,decorations.pathreplacing}
\newcolumntype{P}[1]{>{\centering\arraybackslash}p{#1}}
\title{Studies on the Transcorrelated Method}
\author[1]{Nicholas Lee}
\author[2]{Alex J. W. Thom}
\affil[1]{Physical and Theoretical Chemistry Laboratory, Department of Chemistry, University of Oxford, South Parks Road, Oxford OX1 3QZ, U.K.}
\affil[2]{Yusuf Hamied Department of Chemistry, Lensfield Road, Cambridge, CB2 1EW, U.K.}

\begin{document}
\maketitle

\begin{center}
	\textbf{Abstract}
\end{center}
We investigate the possibility of using a transcorrelated Hamiltonian to describe electron correlation. A method to obtain transcorrelated wavefunctions was developed based on the mathematical framework of the bi-variational principle. This involves the construction of an effective transcorrelated Hamiltonian matrix which can be solved in a self-consistent manner. This was optimised using a method we call Second Order Moment (SOM) minimisation to give highly accurate energies for some closed-shell atoms and helium-like ions. The effect of certain correlator terms on the description of electron-electron and electron-nuclear cusps were also examined graphically and some transcorrelated wavefunctions were compared against near-exact Hylleraas wavefunctions.

\begin{multicols*}{2}

\section{Introduction}
Capturing the effects of electron correlation is a central problem in electronic structure theory. A possible approach to tackle the problem involves the use of a similarity transformed Hamiltonian $\bar{H} = e^{-\tau} \hat{H} e^{\tau}$, where the $\tau$ is a polynomial dependent on electronic positions and incorporates explicitly the correlation between various electron pairs. The use of such a Hamiltonian is known as the \emph{transcorrelated method}. The inclusion of $r_{12}$ terms to describe electronic correlation can be dated back to Hylleraas\autocite{hylleraas_schrodinger_1964}, and later popularised by Kutzelnigg\autocite{kutzelnigg_r12-dependent_1985}, forming the basis of R12/F12 methodology\autocite{liguo_explicitly_2011, ten-no_explicitly_2012} today. Boys and Handy employed the transcorrelated formalism to introduce correlation terms to get near-exact energies for various atoms and molecules\autocite{boys_condition_1969, boys_determination_1969, boys_first_1969, boys_calculation_1969, handy_energies_1969, handy_transcorrelated_1971, handy_towards_1973}. This was done using a custom basis set and an optimised Jastrow factor. Hirschfelder\autocite{hirschfelder_removal_1963}, Bartlett and Nooijen\autocite{nooijen_elimination_1998}, and Klopper and coworkers\autocite{klopper_r12_2000, zweistra_similarity-transformed_2003} have also considered the use of such similarity-transformed Hamiltonians to eliminate the singularities associated with the $\frac{1}{r_{ij}}$ term in the many-electron Hamiltonian.

There are two principal difficulties working with the transcorrelated Hamiltonian. Firstly, the transcorrelated Hamiltonian will involve three-electron operators which can be expensive computationally. Secondly, the transcorrelated Hamiltonian is non-Hermitian. Unlike with Hermitian operators, the variational principle does not hold for non-Hermitian operators. This implies that the expectation value of the transcorrelated Hamiltonian is not bounded from below and hence unphysical energies may be obtained. Furthermore, non-Hermitian matrices are difficult to work with due to the possibility for numerical instability in matrix computations.

However, with the introduction of Variational Monte Carlo (VMC), the calculation became more computationally feasible and promising results were shown for a variety of atoms, molecules\autocite{zweistra_similarity-transformed_2003, umezawa_transcorrelated_2003, umezawa_practical_2005, umezawa_role_2006} and periodic systems\autocite{ochi_optimization_2013, ochi_optical_2014}. To tackle the issue of non-Hermiticity, Luo proposed to replace the non-Hermitian ansatz with a Hermitian approximation so that a variational approach becomes viable\autocite{luo_variational_2010, luo_complete_2011}. 

The transcorrelated Hamiltonian has also more recently been used with a variety of quantum chemistry methods with promising results. Alavi and co-workers have applied Full Configuration Interaction Quantum Monte Carlo (FCIQMC) to the transcorrelated ansatz for a variety of systems successfully\autocite{luo_combining_2018, cohen_similarity_2019, dobrautz_compact_2019, jeszenszki_eliminating_2020}. Their numerical results show that the unboundedness of the non-Hermitian operator did not pose serious difficulties and that highly accurate results, even up to spectroscopic accuracies\autocite{guther_binding_2021}, could be realised. More recently, they have used the transcorrelated Hamiltonian with Coupled Cluster\autocite{schraivogel_transcorrelated_2021} and the energies found demonstrated better basis set convergence. On the other hand, Reiher and co-workers have developed a transcorrelated analogue of Density Matrix Renormalisation Group (DMRG) and applied it to the Fermi-Hubbard model and some homo-nuclear diatomics\autocite{baiardi_transcorrelated_2020, baiardi_explicitly_2022}. They have similarly found that the use of transcorrelation accelerates the convergence to the complete basis set limit.

Building on the recent successes of the transcorrelated method, this work shall attempt a deterministic approach and examine its viability as a computational tool to capture the effects of electron correlation. While Boys and Handy's formulation was deterministic, the computational resources of their time would have restricted the scope of their work. We therefore think that it would be useful to explore the possibilities of a deterministic approach with the computational resources today. Unlike Boys and Handy, however, we will solve the non-Hermitian Hamiltonian matrix self-consistently. Instead of parametrising the correlator using the contraction equations, we propose to find it via what we shall call second-order-moment (SOM) minimisation, an analogue of variance minimisation that we have adapted for this work.\\

\section{Theoretical Background}
\subsection{Transcorrelated Hamiltonian}
The transcorrelated Hamiltonian, $\bar{H} = e^{-\tau} \hat{H} e^{\tau}$, can be expanded using the Baker--Campbell--Hausdorff (BCH) expansion:
\begin{align}
\bar{H}  = \hat{H} + [\hat{H}, \tau] + \frac{1}{2!}[[\hat{H}, \tau], \tau] + \frac{1}{3!}[[[\hat{H}, \tau], \tau], \tau] + ...
\label{TC: BCH expansion}
\end{align}
By using a correlator of the form $\tau = \sum_{i<j} u(\boldsymbol{r}_{i}, \boldsymbol{r}_{j})$, the third- and higher-order commutator terms vanishes. The commutators can be further expanded to give:
\begin{align}
\bar{H}  = \hat{H} - \sum_{i}^{N_{e}} \Big(\frac{1}{2} \nabla_{i}^{2} \tau + \boldsymbol{\nabla}_{i} \tau \cdot \boldsymbol{\nabla}_{i} + \frac{1}{2} \Big(\boldsymbol{\nabla}_{i}\tau \Big)^{2} \Big)
\label{TC: TC expanded}
\end{align}
where $N_{e}$ is the total number of electrons in the system studied.\\
Substituting $\tau = \sum_{i<j} u(\boldsymbol{r}_{i}, \boldsymbol{r}_{j})$, the transcorrelated Hamiltonian takes the form
\begin{equation}
	\bar{H}  = \hat{H} - \sum_{i<j}^{N_{e}} \hat{K}(\boldsymbol{r}_{i}, \boldsymbol{r}_{j}) - \sum_{i<j<k}^{N_{e}} \hat{L}(\boldsymbol{r}_{i}, \boldsymbol{r}_{j}, \boldsymbol{r}_{k})
\label{TC: Transcorrelated equation}
\end{equation}
where
\begin{equation}
\begin{aligned}
\hat{K}(\boldsymbol{r}_{i}, \boldsymbol{r}_{j}) &= \frac{1}{2} \Big( \nabla^{2}_{i} u(\boldsymbol{r}_{i}, \boldsymbol{r}_{j}) + \nabla^{2}_{j} u(\boldsymbol{r}_{i}, \boldsymbol{r}_{j}) \\
&\quad + (\boldsymbol{\nabla}_{i} u(\boldsymbol{r}_{i}, \boldsymbol{r}_{j})^{2} + (\boldsymbol{\nabla}_{j} u(\boldsymbol{r}_{i}, \boldsymbol{r}_{j})^{2} \Big) \\
&\quad + \boldsymbol{\nabla}_{i} u(\boldsymbol{r}_{i}, \boldsymbol{r}_{j}) \cdot \boldsymbol{\nabla}_{i} + \boldsymbol{\nabla}_{j} u(\boldsymbol{r}_{i}, \boldsymbol{r}_{j}) \cdot \boldsymbol{\nabla}_{j} 
\end{aligned}
\end{equation}
\begin{equation}
\begin{aligned}
\hat{L}(\boldsymbol{r}_{i}, \boldsymbol{r}_{j}, \boldsymbol{r}_{k}) &= \boldsymbol{\nabla}_{i} u(\boldsymbol{r}_{i}, \boldsymbol{r}_{j}) \cdot \boldsymbol{\nabla}_{i} u(\boldsymbol{r}_{i}, \boldsymbol{r}_{k}) \\
&\quad + \boldsymbol{\nabla}_{j} u(\boldsymbol{r}_{j}, \boldsymbol{r}_{k}) \cdot \boldsymbol{\nabla}_{j} u(\boldsymbol{r}_{j}, \boldsymbol{r}_{i}) \\
&\quad + \boldsymbol{\nabla}_{k} u(\boldsymbol{r}_{k}, \boldsymbol{r}_{i}) \cdot \boldsymbol{\nabla}_{k} u(\boldsymbol{r}_{k}, \boldsymbol{r}_{j})
\end{aligned}
\end{equation}
The presence of the terms $\boldsymbol{\nabla}_{i} u(\boldsymbol{r}_{i}, \boldsymbol{r}_{j}) \cdot \boldsymbol{\nabla}_{i} + \boldsymbol{\nabla}_{j} u(\boldsymbol{r}_{i}, \boldsymbol{r}_{j}) \cdot \boldsymbol{\nabla}_{j} $ in $\hat{K}(\boldsymbol{r}_{i}, \boldsymbol{r}_{j})$ makes the transcorrelated Hamiltonian non-self-adjoint. This has been derived previously in several papers\autocite{boys_determination_1969, klopper_r12_2000, zweistra_similarity-transformed_2003, cohen_similarity_2019}, but is recapitulated here for completeness. \\

\subsection{One-electron effective Hamiltonian}
The transcorrelated Hamiltonian is non-self-adjoint and will therefore have left- and right-eigenvectors. The left- and right- eigenvectors $\Psi = \hat{\mathcal{A}} (\psi_{1} \psi_{2} \cdots \psi_{n})$ and $\Phi = \hat{\mathcal{A}} (\phi_{1} \phi_{2} \cdots \phi_{n})$ are Slater Determinants formed from molecular orbitals $\{ \psi_{1} \psi_{2} \cdots \psi_{n} \}$ and $\{ \phi_{1} \phi_{2} \cdots \phi_{n} \}$, respectively. A bi-orthonormal set of molecular orbitals, that is, $\braket{ \psi_{i} | \phi_{j} } = \delta_{ij}$ can always be found via Löwdin pairing\autocite{mayer_simple_2003} and hence we assume bi-orthonormality throughout this paper.

The Slater Determinants satisfy the following equations:
\begin{align}
\bar{H} \Phi &= E \Phi	&	\Psi \bar{H}  &= E \Psi
\end{align}
with $E$ denoting the energy associated with the eigenvectors.
The transcorrelated energy can be identified as:
\begin{align}
E &= \frac{ \braket{ \Psi | \bar{H} | \Phi } }{ \braket{ \Psi | \Phi } } \\
	&= \braket{ \Psi | \bar{H} | \Phi }
\end{align}
The denominator is unity due to the bi-orthonormality condition. The effective transcorrelated Hamiltonian can be found by taking the functional variation of the transcorrelated energy. The functional variation can be found by using the method of Lagrange multipliers. Forming the Lagrangian $\mathcal{L}$ under the constraint of bi-orthonormal orbitals:
\begin{equation}
\mathcal{L} = E - \sum_{i=1}\sum_{j=1} \epsilon_{ij} (\braket{ \psi_{i} | \phi_{j} } -\delta_{ij} )
\end{equation}
We seek the solution to $\delta \mathcal{L}$ to find a stationary point of the energy with respect to the constraint. We prove in Appendix \ref{Appendix A} that using the condition $\delta \mathcal{L} = 0$, we get the equation:
\begin{equation}
\begin{split}
\Big[ \hspace{0.2em} \hat{h}  + \sum_{j=1} \bar{G}_{j} + \frac{1}{2}\sum_{j=1}\sum_{k=1} \bar{L}_{jk} \hspace{0.2em} \Big] \phi_{i}(\boldsymbol{r}_{1}) &= \sum_{j=1} \epsilon_{ij} \phi_{i}(\boldsymbol{r}_{1})\\
\bar{H}_{\text{eff}}(\boldsymbol{r}_{1}) \phi_{i}(\boldsymbol{r}_{1}) &= \sum_{j=1} \epsilon_{ij} \phi_{i}(\boldsymbol{r}_{1})\\
\end{split}
\label{TC: Effective Hamiltonian}
\end{equation}
such that
\end{multicols*}

\begin{equation}
\bar{G}_{j}  = \sum_{j=1} \int d\boldsymbol{r}_{2} \psi_{j}^{*}(\boldsymbol{r}_{2} ) ( r_{12}^{-1} -  \hat{K}(\boldsymbol{r}_{1}, \boldsymbol{r}_{2}) ) \mathcal{P}_{2} \phi_{j}(\boldsymbol{r}_{2} )
\end{equation}
\begin{equation}
\bar{L}_{jk} = \int \int d\boldsymbol{r}_{2} d\boldsymbol{r}_{3} \psi_{j}^{*}(\boldsymbol{r}_{2})\psi_{k}^{*}(\boldsymbol{r}_{3})\hat{L}(\boldsymbol{r}_{1}, \boldsymbol{r}_{2}, \boldsymbol{r}_{3}) \mathcal{P}_{3} \phi_{j}(\boldsymbol{r}_{2})\phi_{k}(\boldsymbol{r}_{3}) 
\end{equation}

\begin{multicols}{2}

We also introduce a notation $\mathcal{P}_{N} = \sum_{\hat{P} \in S_{N}} (-1)^{p} \hat{P}$. $S_{N}$ is the symmetric group of degree $N$. For example,
\begin{equation}
\begin{split}
	\mathcal{P}_{3} \ket{ i j k } = \ket{ i j k } - \ket{ i k j } + \ket{ j k i } - \ket{ j i k } + \ket{ k i j } - \ket{ k j i }
\end{split}
\label{TC: Permutations}
\end{equation}
$\mathcal{P}_{3}$ therefore gives all the possible permutations (with the correct parity) of the three-particle ket $\ket{ i j k }$.
$\bar{H}_{\text{eff}}$ is the effective transcorrelated Hamiltonian. It is a functional of the bi-orthogonal set of molecular orbitals $\{\psi_{i} \}$ and $\{\phi_{i} \}$ and can thus be solved for iteratively through a self-consistent approach.\\

\subsection{Jastrow factor}
The following form of the correlator was first introduced by Boys and Handy\autocite{boys_calculation_1969}:
\begin{equation}
u(\boldsymbol{r}_{i}, \boldsymbol{r}_{j}) = \sum_{mno} c_{mno} \Delta_{mn} (\bar{r}_{iA}^{m}\bar{r}_{jA}^{n} + \bar{r}_{iA}^{n}\bar{r}_{jA}^{m})\bar{r}_{ij}^{o}
\label{JF: Jastrow factor}
\end{equation}
where
\begin{equation}
\bar{r} = \frac{ar}{1 + br}
\end{equation}
and
\begin{equation}
\Delta_{mn} = \begin{cases} 
      						\frac{1}{2} & m = n \\
      						1 & \text{otherwise}
   						\end{cases}
\end{equation}
Scaling of the inter-particle distances as $\bar{r}$ is known as the Pad\'e form\autocite{umrigar_optimized_1988}. Scaled distances are commonly used for Jastrow factors such that at large inter-particle distances, the terms in the Jastrow factors will approach a constant. There have been a number of scaling functions employed in literature\autocite{austin_quantum_2012}. Following the work of Schmidt and Moskowitz\autocite{moskowitz_correlated_1992, moskowitz_monte_1982}, we will use the Pad\'e form with $a = b = 1$ due to the simplicity of implementation.

\subsection{Optimising correlator parameters}
The correlators are a function of the set of parameters $\{ c_{mno} \}$. However, determination of these parameters is a non-trivial task. While the parameter $c_{001}$ in equation \ref{JF: Jastrow factor} has been determined previously to be $\frac{1}{2}$ to satisfy the cusp condition, the other parameters $c_{mno}$ have yet to be determined. The unbounded nature of the non-self-adjoint transcorrelated Hamiltonian operator prevents the use of energy minimisation for this. However, minimisation of the local energy variance can be performed to find these parameters. Schmidt and Moskowitz applied Variational Monte Carlo (VMC) to calculate and minimise the variance. They performed this with correlators consisting of 7, 9, and 17 terms and found that with a 17 term correlator, 68--100\% of the correlation energies for atoms helium through neon could be recovered using their variance minimised parameters.\\
Handy also independently developed a variance minimisation procedure to optimise the transcorrelated parameters\autocite{handy_minimzation_1971}. He introduced the transcorrelated variance:
\begin{equation}
	U^{\text{TC}} = \braket{ e^{-\tau} \hat{H} e^{\tau} \Phi | e^{-\tau} \hat{H} e^{\tau} \Phi } - \braket{ \Phi | e^{-\tau} \hat{H} e^{\tau} | \Phi }^{2}
\label{JF: Handy's variance}
\end{equation}
The minimisation of $U^{TC}$ was performed through the Davidson method and near-exact energy for the helium atom was calculated through this method, albeit with a slight modification of the Jastrow factor. However, helium is a two-electron system; for any systems with more than two electrons, the three-electron operator in the transcorrelated Hamiltonian will in general give a non-zero term. As such, the calculation of the transcorrelated variance in equation \ref{JF: Handy's variance} will require the evaluation of six-electron operators. The high computational cost and poor scaling has deterred research efforts along this line of inquiry.

\subsection{Second-Order-Moment (SOM) Minimisation}
While variance is well-defined for a self-adjoint operator, there is little literature for its non-self-adjoint counterpart.
It is well known from Linear Algebra that a non-Hermitian matrix has left- and right-eigenvectors which are not necessarily identical. The transcorrelated Hamiltonian is a non-self-adjoint operator and would similarly have left- and right-eigenfunctions $\Psi$ and $\Phi$ respectively. We assume the use of a bi-orthogonal basis such that $\braket{\Psi | \Phi} = 1$. Instead of variance minimisation, we propose the minimisation of the second-order-moment (SOM), a bi-orthogonal analogue of the variance for a non-self-adjoint Hamiltonian:
\begin{equation}
U^{\text{SOM}} = \braket{ \Psi | ( \bar{H} - \langle \bar{H} \rangle ) ( \bar{H} - \langle \bar{H} \rangle ) | \Phi }
\end{equation}
where $\langle \bar{H} \rangle = \braket{ \Psi | \bar{H} | \Phi }$.
This is a bi-orthogonal extension to the usual definition of the variance (or second central moment in some papers\autocite{small_correction_2007, small_central_2008}). To the best of the authors' knowledge, the minimisation of $U^{\text{SOM}}$ has not previously been performed. We shall first analyse some limiting cases to gain a better understanding of the quantity $U^{\text{SOM}}$.\\

In the limit of $\bar{H}^{\dagger}  = \bar{H}$ (self-adjointness), the left- and right-eigenfunctions become identical, $\Psi = \Phi$. $U^{\text{SOM}}$ therefore reduces to the standard definition of the variance:
\begin{equation}
\begin{split}
U^{\text{SOM}} &= \braket{ \Psi | \bar{H} \bar{H} | \Phi } - \langle \bar{H} \rangle^{2} \\
						 &= \braket{ \Phi | \bar{H}^{\dagger} \bar{H} | \Phi } - \langle \bar{H} \rangle^{2} \\
						 &=\braket{ \Psi |  \bar{H}^{2} | \Phi } - \braket{ \Phi |  \bar{H} | \Phi } ^{2} \\
\end{split}
\end{equation}
In the limit that $\Phi$ is an exact eigenfunction, that is, $\bar{H} \Phi = \lambda \Phi$ where $\lambda \in \mathbb{C}$,
\begin{equation}
\begin{split}
\langle \bar{H} \rangle &= \braket{ \Psi | \bar{H} | \Phi } \\
									 &= \lambda \braket{ \Psi | \Phi } \\
									 & = \lambda
\end{split}
\end{equation}
where we have made use of the bi-orthogonality of the left- and right-eigenfunctions. Then,
\begin{equation}
\begin{split}
U^{\text{SOM}} &= \braket{ \Psi | ( \bar{H} - \lambda ) ( \bar{H} - \lambda ) | \Phi }\\
						 &= ( \lambda - \lambda ) \braket{ \Psi | ( \bar{H} - \lambda ) | \Phi }\\
						 &= 0
\end{split}
\end{equation}
A similar proof holds for the limit that $\Psi$ is an exact eigenfunction. Since the exact eigenfunction has to satisfy the condition that $U^{\text{SOM}} = 0$, the parameters should be varied such that the quantity $U^{\text{SOM}}$ becomes as close to zero as possible. The evaluation of $U^{\text{SOM}}$ would similarly require the evaluation of six-electron terms (three from each $\bar{H}$) and it is therefore as computationally challenging as Handy's transcorrelated variance. To side-step this difficulty, the resolution of identity is employed. In the bi-orthogonal basis, the identity is given by:
\begin{equation}
\mathbb{I} = \sum_{k} \ket{\Phi_{k}} \bra{\Psi_{k}}
\end{equation}
where $k$ runs through all of the possible Slater Determinants for the given basis.

\begin{equation}
\begin{split}
U^{\text{SOM}} &= \braket{ \Psi | (\bar{H} - \langle \bar{H} \rangle ) ( \bar{H} - \langle \bar{H} \rangle ) | \Phi } \\
				  		&= \sum_{k} \braket{ \Psi | ( \bar{H} - \langle \bar{H} \rangle ) | \Phi_{k} } \braket{ \Psi_{k} |(\bar{H} - \langle \bar{H} \rangle ) | \Phi } \\
				  		&= (\braket{ \Psi | \bar{H} | \Phi } - \langle \bar{H} \rangle)(\braket{ \Psi | \bar{H} | \Phi } \\
				  		&\quad - \langle \bar{H} \rangle) +\sum_{k \neq 0} \braket{ \Psi | \bar{H} | \Phi_{k} } \braket{ \Psi_{k} | \bar{H} | \Phi } \\
				  		&= \sum_{k \neq 0} \braket{ \Psi | \bar{H} | \Phi_{k} } \braket{ \Psi_{k} | \bar{H} | \Phi } \\
				  		&\approx \sum_{\sigma} \sum_{ia} \braket{ \Psi | \bar{H} | \Phi_{i}^{a} } \braket{ \Psi_{i}^{a} | \bar{H} | \Phi } \\
				  		&\quad + \frac{1}{2}\sum_{\sigma \sigma'} \sum_{ijab} \braket{ \Psi | \bar{H} | \Phi_{ij}^{ab} } \braket{ \Psi_{ij}^{ab} | \bar{H} | \Phi } \\
\end{split}
\end{equation}
where the factor of a half was added to take into account double counting of $ij$ and $ab$. The $\sigma$ terms denote the various spins of electrons. The penultimate step is an approximation as we ignore the triple excitation terms when they are much smaller than the double excitation terms. In addition, the single excitation term has terms with: $\braket{ \Psi_{i}^{a} | \bar{H} | \Phi }$. By analogy to Brillouin's theorem for the Hartree--Fock method, single excitation determinants will not interact \emph{directly} with the ground-state determinant, that is, $\braket{ \Psi_{i}^{a} | \bar{H} | \Phi } = 0$. 
We can therefore ignore the single excitation terms and deduce that:
\begin{equation}
U^{\text{SOM}} \approx \frac{1}{2}\sum_{\sigma \sigma'} \sum_{ijab} \braket{ \Psi | \bar{H} | \Phi_{ij}^{ab} } \braket{ \Psi_{ij}^{ab} | \bar{H} | \Phi }
\end{equation}

\subsection{Bi-variational Principle}
Having found the appropriate correlator parameters, we can construct the transcorrelated Hamiltonian and solve for its eigenfunctions. However, when the Hamiltonian is non-self-adjoint (as in the case for the transcorrelated Hamiltonian), the variational principle does not hold and the expectation value of the Hamiltonian is not bounded from below. A naive minimisation of the Hamiltonian's expectation value can therefore lead to values below the exact ground state energy, which are unphysical. However, one can formulate a different variational principle for a generic operator, which is not necessarily self-adjoint. While the mathematical exposition on the bi-variational principle has been previously undertaken by Löwdin\autocite{lowdin_stability_1983, froelich_hartree--fock_1983, lowdin_linear_1998}, the essential parts of the proofs are reviewed here as it is a crucial to the development of the transcorrelated method.

For a non-self-adjoint operator $\bar{H}$, we can define left and right eigenfunctions $\Psi$ and $\Phi$, respectively such that
\begin{align}
\bar{H}\Phi &= \lambda \Phi & \bar{H}^{\dagger}\Psi &= \mu \Psi & \lambda, \mu \in \mathbb{C}
\label{BV: Adjoint operators}
\end{align}
We note that $\lambda$ and $\mu$ are related by complex conjugation, that is, $\lambda = \mu^{*}$
For a given pair of trial functions $\Psi_{i}$ and $\Phi_{i}$ such that
\begin{align}
\Phi_{i} &= \Phi + \delta \Phi & \Psi_{i} &= \Psi + \delta \Psi
\label{BV: Trial functions}
\end{align}
the expectation values $\lambda_{i}$ and $\mu_{i}$ are given by
\begin{align}
\lambda_{i} &= \lambda + \frac{\braket{ \delta\Psi | \bar{H} - \lambda| \delta\Phi }}{ \braket{\Psi_{i} | \Phi_{i} } } & \mu_{i} &= \mu + \frac{\braket{ \delta\Phi | \bar{H}^{\dagger} - \mu| \delta\Psi }}{ \braket{\Phi_{i} | \Psi_{i} } } 
\label{BV: Second order error I}
\end{align}
The expectation values $\lambda_{i}$ and $\mu_{i}$ have vanishing first-order variations ($\delta \lambda_{i} = 0$), that is, they correspond to stationary points about the exact eigenvalues $\lambda$ and $\mu$, respectively. This is known as the \emph{bi-variational principle for a pair of adjoint operators}\autocite{lowdin_stability_1983}.

Conversely, we can show that if $\delta \lambda_{i} = 0$ for all $\delta \Phi$ and $\delta \Psi$,
\begin{align}
	(\bar{H} - \lambda_{i}) \Phi_{i} &= 0 & (\bar{H} - \lambda_{i})^{\dagger} \Psi_{i} &= 0 
\label{BV: Second order error II}
\end{align}
This implies that the trial function $\Phi_{i}$ is an eigenfunction of $\bar{H}$ with eigenvalue $\lambda_{i}$ and  $\Psi_{i}$ is an eigenfunction of $\bar{H}^{\dagger}$ with eigenvalue $\lambda_{i}^{*} = \mu_{i}$. Equation \ref{BV: Second order error I} implies that if the trial functions $\Phi_{i}$ and $\Psi_{i}$ are correct to first order, the approximation of the eigenvalue $\lambda_{i}$ to the exact eigenvalue $\lambda$ is correct to second order. 

\subsection{Matrix Representation}
The bi-variational equations can be recast in matrix form. In the following, the tensor notation of Head-Gordon \emph{et al}.\autocite{head-gordon_tensor_1998} shall be used. Given an atomic-orbital basis $\{ \chi_{1} ... \chi_{n}\}$, we can expand any pair of trial functions $\Phi_{i}$ and $\Psi_{i}$ as
 \begin{align}
\Phi_{i} &= \sum_{\tau} \chi_{\tau} ?[l]c^{\tau}_{i}? & \Psi_{i} &= \sum_{\tau} \chi_{\tau} ?[l]d^{\tau}_{i}?
\end{align}
From equation \ref{BV: Second order error II}, the bi-variation equations can then be expressed as
\begin{equation}
\begin{split}
\bar{H} \Phi_{i}  &= \lambda_{i} \Phi_{i} 	\\
\bar{H}^{\dagger} \Psi_{i}  &= \lambda_{i}^{*} \Psi_{i}
\end{split}
\end{equation}
\begin{equation}
\begin{split}
\sum_{\tau}\braket{ \chi_{\sigma} | \bar{H} | \chi_{\tau} } ?[l]c^{\tau}_{i}? &= \lambda \sum_{\tau} \braket{ \chi_{\sigma} | \chi_{\tau} } ?[l]c^{\tau}_{i}? \\
\sum_{\tau}\braket{ \chi_{\sigma} | \bar{H}^{\dagger} | \chi_{\tau} } ?[l]d^{\tau}_{i}? &= \lambda^{*} \sum_{\tau} \braket{ \chi_{\sigma} | \chi_{\tau} } ?[l]d^{\tau}_{i}? 
\end{split}
\label{BV: MatrixEquations}
\end{equation}
\begin{equation}
\begin{split}
\boldsymbol{ \bar{H} c} &= \boldsymbol{ \Lambda S c} \\
\boldsymbol{ \bar{H}^{\dagger} d} &= \boldsymbol{ \Lambda^{\dagger} S d}
\end{split}
\end{equation}
The expressions in equation \ref{BV: MatrixEquations} were obtained through left-multiplying by $\chi_{\sigma}$ and integrating over all space. In the last step we make the identification that $\boldsymbol{ \bar{H} }_{\sigma \tau} = \braket{ \chi_{\sigma} | \bar{H}^{\dagger} | \chi_{\tau} } $ and $\boldsymbol{S}_{\sigma \tau} = \braket{ \chi_{\sigma} | \chi_{\tau} }$.

\subsection{Solving the Transcorrelated Equation}
We are now in a position to apply the bi-variational approach on the transcorrelated Hamiltonian. The effective transcorrelated Hamiltonian matrix has to be solved iteratively as the two- and three-electron terms are dependent the trial functions $\Phi_{i}$ and $\Psi_{i}$. The following workflow was utilised:
\begin{enumerate}
\item Perform Hartree--Fock calculation and use the Hartree--Fock coefficients as a starting guess.
\item Build the effective transcorrelated Hamiltonian matrix.
\item Diagonalise the matrix to get new coefficients for the left- and right-eigenvectors.
\item Repeat until convergence.
\end{enumerate}
In doing so, we are \emph{simultaneously} optimising both the left- and right-eigenvectors. This is a different approach to that of Dobrautz, Luo, and Alavi\autocite{dobrautz_compact_2019} where only the right-eigenvector is optimised. While our approach requires the optimisation of both left- and right-eigenvectors, which translates to a more expensive calculation, we gain the benefit of bounding the error of the calculation by the bi-variational principle.\\

\subsection{Maximum Overlap Method}
Convergence of the bi-variational approach can be difficult in some cases. Taking inspiration from the work of Gilbert and co-workers\autocite{gilbert_self-consistent_2008}, we first assume that the Hartree--Fock coefficients are a good guess at our final coefficients. Therefore, at each iteration, the set of orbitals with the largest overlap to the occupied orbitals in the previous iterations will be picked. This process proceeds until convergence is reached. This is known as the Maximum Overlap Method (MOM). Given the right coefficient matrix from the previous iteration $\boldsymbol{C}_{\text{old}}$, the left coefficient matrix from the current iteration $\boldsymbol{D}_{\text{new}}$ and the atomic orbital overlap matrix $\boldsymbol{S}$, the maximum overlap matrix $\boldsymbol{O}_{\text{MOM}}$ is given by:
\begin{align}
	\boldsymbol{O}_{\text{MOM}} = | \boldsymbol{C}^{\dagger}_{\text{old}} \boldsymbol{S} \boldsymbol{D}_{\text{new}} |
\end{align}
The bi-orthogonal solutions from each iteration are determined only up to a phase factor, and hence the modulus is taken to ensure that the overlap remains positive.

Even with traditional implementations of MOM, it is found that it is possible for SCF iterations to converge onto unwanted solutions. This has led to the introduction of the Initial Maximum Overlap Method (IMOM)\autocite{barca_simple_2018}, where new orbitals in each iteration are picked based on their overlaps with the initial guess orbitals. This prevents the solutions from drifting away from the initial guess and has been shown to give better convergence to desired solutions. In this work, we adapt it for bi-orthogonal orbitals, such that the maximum overlap matrix $\boldsymbol{O}_{\text{IMOM}}$ is given by:
\begin{align}
	\boldsymbol{O}_{\text{IMOM}} = | \boldsymbol{C}^{\dagger}_{\text{initial}} \boldsymbol{S} \boldsymbol{D}_{\text{new}} |
\end{align}
where $\boldsymbol{C}_{\text{initial}}$ is the initial left coefficient matrix.\\
Both forms of the maximum overlap method were implemented for improved convergence of the iterative procedure.\\

\section{Computational Details}
The transcorrelated method is implemented in Python. The matrix elements relating to the correlator were found via numerical integration. These integrations were performed with grids found in the PySCF package. Throughout this work, we used Treutler--Ahlrichs grids with Becke partitioning. Q-Chem 5.3 was used for conventional NOCI calculations and for finding Hartree--Fock solutions. Mathematica was used to plot Figures \ref{Angular}-\ref{TCRadial}.

\section{Transcorrelated energies using Schmidt--Moskowitz Parameters}
\begin{table*}
\begin{tabular}{c c}
\hline
\hline
Set			&		Parameters\\
\hline
SM7			&		001, 002, 003, 004, 200, 300, 400 \\
SM9			&		001, 002, 003, 004, 200, 300, 400, 220, 202\\
SM17		&		001, 002, 003, 004, 200, 300, 400, 220, 202, 222, 402, 204, 422, 602, 404, 224, 206\\
SOM8		&		001, 002, 003, 004, 100, 200, 300, 400 \\
SOM10		&		001, 002, 003, 004, 100, 200, 300, 400, 220, 202\\
SOM18		&		001, 002, 003, 004, 100, 200, 300, 400, 220, 202, 222, 402, 204, 422, 602, 404, 224, 206\\
\hline
\hline
\end{tabular}
\caption{Summary of the various sets of parameters used, where each number in the second column has the form $mno$. For example, $001$ corresponds to the $m=0$, $n=0$, and $o=1$ term. i.e. $r_{ij}$ term. "SM" refers to Schmidt--Moskowitz parameters while "SOM" refers to parameters found via SOM minimisation.}
\label{Params}
\end{table*}

\begin{table*}
\centering
\begin{tabular}{c c c c c c c}
\hline
\hline
		&			cc-pVDZ			&			cc-pVTZ			&			cc-pVQZ			&			SM7\autocite{schmidt_correlated_1998}	&			FCIQMC (cc-pVQZ)\autocite{cohen_similarity_2019}		&			Experimental\autocite{chakravorty_ground-state_1993}	\\
\hline
He	&			-2.8962			&			-2.9021			&			-2.9025			&			-2.8997			&			-									&			-2.9037					\\
Li		&			-7.4670			&			-7.4671			&			-7.4672			&			-7.4746			&			-7.4779						&			-7.4781					\\
Be		&			-14.6111			&			-14.6112			&			-14.6113			&			-14.6259			&			-14.6679						&			-14.6674					\\
B		&			-24.5740			&			-24.5756			&			-24.5764			&			-24.5946			&			-24.65417					&			-24.6539					\\
C		&			-37.7431			&			-37.7475			&			-37.7489			&			-37.7721			&			-37.8479						&			-37.8450					\\
N		&			-54.4502			&			-54.4593			&			-54.4618			&			-54.5019			&			-54.5878						&			-54.5892					\\
O		&			-74.8659			&			-74.8849			&			-74.8659			&			-74.9469			&			-75.0630						&			-75.0673					\\
F		&			-99.4619			&			-99.4912			&			-99.4989			&			-99.5746			&			-99.7251						&			-99.7339					\\
Ne	&			-128.6119		&			-128.6528		&			-128.6640		&			-128.7689		&			-128.9297					&			-128.9376				\\
\hline
\hline
\end{tabular}
\caption{Comparison of the transcorrelated total energies (in Hartrees) found with the bi-variational approach using 7 parameters against literature and experimental values. The parameters were the same as that used by Schmidt and Moskowitz\autocite{schmidt_correlated_1998}. FCIQMC (cc-pVQZ basis) data was found by Alavi and co-workers\autocite{cohen_similarity_2019}. Experimental values were found by Chakravorty and co-workers\autocite{chakravorty_ground-state_1993}.}
\label{7}
\end{table*}

\begin{table*}
\centering
\begin{tabular}{c c c c c c c}
\hline
\hline
		&			cc-pVDZ			&			cc-pVTZ			&			cc-pVQZ			&			SM9\autocite{schmidt_correlated_1998}					&			FCIQMC\autocite{cohen_similarity_2019}		&			Experimental\autocite{chakravorty_ground-state_1993}			\\
\hline
He	&			-2.8935			&			-2.8995			&			-2.8998			&			-2.9029			&			-				&			-2.9037					\\
Li		&			-7.4746			&			-7.4727			&			-7.4724			&			-7.4731			&			-				&			-7.4781					\\
Be		&			-14.6205			&			-14.6191			&			-14.6192			&			-14.6332			&			-				&			-14.6674					\\
B		&			-24.6057			&			-24.6055			&			-24.6062			&			-24.6113			&			-				&			-24.6539					\\
C		&			-37.7592			&			-37.7632			&			-37.7644			&			-37.7956			&			-				&			-37.8450					\\
N		&			-54.5262			&			-54.5334			&			-54.5349			&			-54.5390			&			-				&			-54.5892					\\
O		&			-74.9971			&			-75.0136			&			-75.0164			&			-75.0109			&			-				&			-75.0673					\\
F		&			-99.6589			&			-99.6873			&			-99.6920			&			-99.6685			&			-				&			-99.7339					\\
Ne	&			-128.8567		&			-128.8985		&			-128.9070		&			-128.8796		&			-				&			-128.9376				\\
\hline
\hline
\end{tabular}
\caption{Comparison of the transcorrelated total energies (in Hartrees) found with the bi-variational approach using 9 parameters against literature and experimental values. The parameters were the same as that used by Schmidt and Moskowitz\autocite{schmidt_correlated_1998}. FCIQMC (cc-pVQZ basis) data was found by Alavi and co-workers\autocite{cohen_similarity_2019}. Experimental values were found by Chakravorty and co-workers\autocite{chakravorty_ground-state_1993}.}
\label{9}
\end{table*}

\begin{table*}
\centering
\begin{tabular}{c c c c c c c}
\hline
\hline
		&			cc-pVDZ			&			cc-pVTZ			&			cc-pVQZ			&			SM17\autocite{schmidt_correlated_1998}					&			FCIQMC\autocite{cohen_similarity_2019}					&			Experimental\autocite{chakravorty_ground-state_1993}			\\
\hline
He	&			-2.8959			&			-2.9020			&			-2.9023			&			-2.9036			&			-							&			-2.9037					\\
Li		&			-7.4770			&			-7.4766			&			-7.4765			&			-7.4768			&			-7.4785				&			-7.4781					\\
Be		&			-14.6304			&			-14.6300			&			-14.6283			&			-14.6370			&			-14.6675				&			-14.6674					\\
B		&			-24.5974			&			-24.5980			&			-24.5977			&			-24.6156			&			-24.6529				&			-24.6539					\\
C		&			-37.7740			&			-37.7766			&			-37.7772			&			-37.8017			&			-37.8446				&			-37.8450					\\
N		&			-54.5022			&			-54.5099			&			-54.5116			&			-54.5456			&			-54.5884				&			-54.5892					\\
O		&			-74.9549			&			-74.9719			&			-74.9549			&			-75.0146			&			-75.0661				&			-75.0673					\\
F		&			-99.5830			&			-99.6117			&			-99.6187			&			-99.6736			&			-99.7328				&			-99.7339					\\
Ne	&			-128.7533		&			-128.7925		&			-128.8043		&			-128.8796		&			-128.9354			&			-128.9376				\\
\hline
\hline
\end{tabular}
\caption{Comparison of the transcorrelated total energies (in Hartrees) found with the bi-variational approach using 17 parameters against literature and experimental values. The parameters were the same as that used by Schmidt and Moskowitz\autocite{schmidt_correlated_1998}. FCIQMC (cc-pVQZ basis) data was found by Alavi and co-workers\autocite{cohen_similarity_2019}. Experimental values were found by Chakravorty and co-workers\autocite{chakravorty_ground-state_1993}.}
\label{17}
\end{table*}

Schmidt and Moskowitz have previously found sets of 7, 9, and 17 correlator parameters for first-row atoms via variance minimisation \autocite{schmidt_correlated_1998}. Following Alavi and co-workers, we shall refer to these sets as SM7, SM9, and SM17, respectively. The correlator has the form in equation \ref{JF: Jastrow factor}. For ease of reference, the various terms incorporated in the correlator for SM7, SM9, and SM17 are tabulated in Table \ref{Params}.

Using the correlator parameters found by Schmidt and Moskowitz, we solved the transcorrelated Hamiltonian for the first-row atoms with a series of correlation-consistent basis sets (cc-pVXZ, X=D, T, Q). This was done by using the corresponding Unrestricted Hartree--Fock orbitals as a starting guess and varying it until self-consistency. The data in Tables \ref{7}, \ref{9} and \ref{17} shows that for a fixed set of correlator parameters, the transcorrelated total energies of small atoms converge with basis set size. However, the transcorrelated energies do not necessarily decrease with an increasing number of parameters used. For example, in atoms from boron through neon, the transcorrelated energies increase going from 9 parameters to 17 parameters. This can be understood when we consider the origin of the parameters used. Schmidt and Moskowitz used Slater-type orbitals (STOs) in their optimisation studies to obtain the parameters. On the other hand, this work employs Gaussian-type orbitals (GTOs). One major difference between the STOs and GTOs is that the electron-nuclear cusp condition is fulfilled while using STOs but not when using GTOs. Different corrections are therefore required for the Hartree--Fock solutions expressed with different orbital bases, leading to the need for different parameters. Hence, the parameters used in this study may not be optimal.

Alavi and co-workers have also used correlation consistent bases in their work on the transcorrelated Hamiltonian. However, they are able to find energies in excellent agreement with experimental values (Tables \ref{7}, \ref{9} and \ref{17}). We believe that this is due to the effective multi-reference nature of the FCIQMC method such that any errors incurred from using these SM parameters are corrected for by adjusting the weight of each determinant.

\section{SOM minimisation of correlator parameters}

\subsection{Singlet state atoms}
To improve upon the accuracy of our results, we allowed correlator parameters to vary alongside the orbitals. The correlator parameters were optimised by using SOM minimisation. The parameters found from Schmidt and Moskowitz (the set of 7 parameters) were used as a starting guess for the optimisation, with an additional $mno=001$ term to correct for the electron-nuclear cusp conditions (SOM8 in Table \ref{Params}). The starting guess for the $mno=001$ term is zero. This set of 18 parameters will be referred to as SOM18. In practice, we have found it to be useful to optimise the parameters using a two-step SOM minimisation procedure where we first keep the orbitals fixed through the optimisation process and after the first round of optimisation, we perform SOM minimisation with orbital relaxation in each iteration of the second optimisation cycle. In doing so, we are less likely to get caught in local minima after orbital relaxation. The transcorrelated energies found using the optimised parameters are tabulated in Table \ref{SOM-CS}.

\begin{table*}
\centering
\begin{tabular}{c c c c c c}
\hline
\hline
								&			HF					&			SOM8					&			Exact*					&			Difference		& 		Correlation energy (\%)\\
\hline
$^{1}S$ He				&			-2.8615			&		-2.8947					&		-2.9037					&			0.0090				&			79\\
$^{1}S$ Be				&			-14.5730			&		-14.6663					&		-14.6674					&			0.0011				&			99\\
$^{1}S$ C					&			-37.6042			&		-37.7435					&		-37.7465					&			0.0030				&			98\\
$^{1}S$ O					&			-74.6897			&		-74.9093					&		-74.9133					&			0.0040				&			98\\
$^{1}S$ Ne				&			-128.5435		&		-128.8758				&		-128.9376				&			0.0618				&			84\\
\hline
\hline
\end{tabular}
\caption{Comparison of energies found after SOM minimisation and the exact energies. The difference (in Hartrees) and the percentage of correlation energy found were similarly reported. *The exact energies of $^{1}S$ C and $^{1}S$ O were deduced from spectroscopic measurements\autocite{haris_critically_2017, moore_selected_1976}. The exact energies of the other closed shell atoms were taken from experimental values found by found by Chakravorty and co-workers\autocite{chakravorty_ground-state_1993}. The optimised parameters can be found in the Appendix (Table \ref{SOM-CS-Appendix})}
\label{SOM-CS}
\end{table*}

\begin{table*}
\centering
\begin{tabular}{c c }
\hline
\hline
Initial guess													&			TC energy		\\
\hline
$c_{001} = 0.5$, $c_{100} = +1$				&			-2.8969			\\
$c_{001} = 0.5$, $c_{100} = -1$					&			-2.8989			\\
$c_{001} = 0.5$, $c_{100} = -2$					&			-2.9037			\\
\hline
\hline
\end{tabular}
\caption{Transcorrelated (TC) energies of helium atom, in Hartrees for different starting guesses. The set of parameters SOM8 was used, with starting guesses of $0$ unless otherwise stated in the first column. Starting from SOM8 with $c_{001} = 0.5$, $c_{100} = -2$ and other parameters zero, a highly accurate energy of helium atom could be found. The optimised parameters can be found in the Appendix (Table \ref{Different guesses-Appendix})}
\label{Results: Different guesses}
\end{table*}

The energies found did not appear to suffer from non-variationality. For the $^{1}S$ states of Be, C and O, very accurate energies could be found which recover more than 98\% of the correlation energy. 
While the results for helium and neon were not as encouraging, we note that a highly accurate energy of $-2.9037 E_{\text{h}}$ for helium could be found by using a different starting guess (Table \ref{Results: Different guesses}). A similarly good result of $-2.9033 E_{\text{h}}$ (see Table \ref{Results: HF-SOM}) can also be found by using the SOM18 set of parameters.

In practice, there were a number of local minima found during optimisations depending on the starting guess and hence a number of possible energies can in principle be found. A sample of possible solutions for helium are tabulated in Table \ref{Results: Different guesses}. This suggests that the starting guess is very important to obtain the right correlator parameters and one should be cautious about the parameters found from such an optimisation.

 In the case of neon, the use of a larger set of parameters (SOM18) gave a non-variational energy of $-129.0019 E_{\text{h}}$. This demonstrates the possibility of obtaining non-variational energies and highlights the potential pitfalls of using SOM minimisation to obtain correlator parameters.

\subsection{Helium-like systems}

\begin{table*}
\centering
\begin{tabular}{c c c c c c c c }
\hline
\hline
											&			Basis					&			HF					&			SOM18				& 				Exact		&				Error 							&			Error 				& 		Correlation 					\\
											&									&									&									& 								&				(HF)								&			(SOM)				&		energy (\%)					\\
\hline
$^{1}S$ H$^{-}$					&		aug-cc-pVQZ		&		-0.4878				&		-0.5231				&		-0.5278			&			0.0400								&			0.0047				&				88						\\
\hline
He									    &      cc-pVQZ				&      -2.8615              &      -2.9033              &      -2.9037 			&          0.0422                           	&			0.0004            &              99                      \\
\hline
$^{1}S$ Li$^{+}$					&		 cc-pCVQZ			&		-7.2364				&		-7.2807				&		-7.2799			&			0.0435								&		  -0.0008				&				101						\\
\hline
$^{1}S$ Be$^{2+}$				&		 cc-pCVQZ			&		-13.6113				&		-13.6558				&		-13.6556			&			0.0443								&		  -0.0002				&				100						\\
\hline
$^{1}S$ B$^{3+}$					&		 cc-pCVQZ			&		-21.9862				&		-22.0298		    	&		-22.0309			&			0.0447								&			0.0011				&				98						\\
\hline
$^{1}S$ C$^{4+}$				&		 cc-pCVQZ			&		-32.3611				&		-32.4038			    &		-32.4062			&			0.0451								&			0.0024				&				95						\\
\hline	
$^{1}S$ N$^{5+}$				&		 cc-pCVQZ			&		-44.7360				&		-44.7779				&		-44.7814			&			0.0454 							&			0.0035				&				92						\\
\hline
$^{1}S$ O$^{6+}$				&		 cc-pCVQZ			&		-59.1110				&		-59.1486				&		-59.1566			&			0.0456								&			0.0080				&				82						\\
\hline
$^{1}S$ F$^{7+}$					&		 cc-pCVQZ			&		-75.4859				&		-75.5214				&		-75.5317			&			0.0458								&			0.0103				&				78						\\
\hline
$^{1}S$ Ne$^{8+}$				&		 cc-pCVQZ			&		-93.8608			    &		-93.8966				&		-93.9068			&			0.0460								&			0.0102				&				78						\\
\hline				
\hline
\end{tabular}
\caption{Comparison of absolute error in the HF energy against the absolute error in the energy found by SOM minimisation. The absolute error for both HF and SOM minimisation methods increases with the magnitude of nuclear charge.  The absolute error found from HF is consistently above that of those found from SOM minimisation. The optimised correlator parameters found are tabulated in the Appendix (Tables \ref{HF-SOM-Appendix1}, \ref{HF-SOM-Appendix2}).}
\label{Results: HF-SOM}
\end{table*}

\begin{table*}
\centering
\begin{tabular}{c c c c c c }
\hline
\hline
											&			Basis					&			HF					&			SOM18				&				Error 							&			Error 				\\
											&									&									&									&				(HF)								&			(SOM)				\\
\hline
$^{1}S$ H$^{-}$					&		 cc-pVQZ				&		-0.4735				&		-0.5078				&			0.0543								&			0.0200				\\
\hline
$^{1}S$ Li$^{+}$					&		 cc-pVQZ				&		-7.2364				&		-7.2872				&			0.0435								&		  -0.0073				\\
\hline
$^{1}S$ Be$^{2+}$				&		 cc-pVQZ				&		-13.6113				&		-13.6652				&			0.0443								&		  -0.0096				\\
\hline
$^{1}S$ B$^{3+}$					&		 cc-pVQZ				&		-21.9862				&		-22.0455	    		&			0.0447								&		  -0.0146				\\
\hline
$^{1}S$ C$^{4+}$				&		 cc-pVQZ				&		-32.3611				&		-32.4231			    &			0.0451								&		  -0.0169				\\
\hline	
$^{1}S$ N$^{5+}$				&		 cc-pVQZ				&		-44.7360				&		-44.8001				&			0.0454 							&		  -0.0187				\\
\hline
$^{1}S$ O$^{6+}$				&		 cc-pVQZ				&		-59.1108				&		-59.1780				&			0.0458								&		  -0.0214				\\
\hline
$^{1}S$ F$^{7+}$					&		 cc-pVQZ				&		-75.4857				&		-75.5548				&			0.0460								&		  -0.0231				\\
\hline
$^{1}S$ Ne$^{8+}$				&		 cc-pVQZ				&		-93.8605			    &		-93.9309				&			0.0463								&		  -0.0241				\\
\hline				
\hline
\end{tabular}
\caption{Comparison of absolute error in the HF energy against those found by SOM minimisation in cc-pVQZ basis. The absolute error from SOM minimisation is significantly higher than those found in Table \ref{Results: HF-SOM}. Most of the transcorrelated energies found are also non-variational.}
\label{Results: ccpVQZ-SOM}
\end{table*}

Encouraged by the possibility of highly accurate energies using SOM minimisation, we examined the approach on a series of helium-like systems (Table \ref{Results: HF-SOM}). To find the correlator parameters for this series, we used the set of 17 parameters from Schmidt and Moskowitz and with an additional $mno=001$ term as a starting guess for the helium atom and performed SOM minimisation to obtain the optimised parameters (SOM18). These optimised parameters were then used as starting guesses for each of these ions.

We found that it was important to use an augmented basis set (aug-cc-pVQZ) for the negatively charged hydride anion as more diffuse functions are required to describe the expanded orbitals. In contrast, a basis optimised for describing core-core correlations (cc-pCVQZ) was found to be useful to describe the contracted orbitals in cations.

For comparison, the same calculations were performed with a cc-pVQZ basis set and the results are tabulated in Table \ref{Results: ccpVQZ-SOM}. The use of cc-pVQZ basis increased the absolute error from SOM minimisation and the transcorrelated energies found were mostly non-variational. This shows that the choice of basis set is imperative to the accuracy of the transcorrelated method.

Using the appropriate basis sets for cations and anions, we found that the absolute error from SOM minimisation is lower than that for HF for each ion (Table \ref{Results: HF-SOM}). From H$^{-}$ through N$^{5+}$, SOM minimisation recovers a large proportion of the correlation energy. From O$^{6+}$ through Ne$^{8+}$, the percentage of correlation energy recovered drops considerably. This is likely due to the highly contracted nature of the 1s orbitals in these highly charged cations and a bigger basis with more contracted basis functions would be required to more accurately describe the electron correlation. However, it is possible that a better starting guess could similarly improve the correlation energy recovered.

It is also gratifying to note that most of the energies found from SOM minimisation using appropriate basis sets do not exhibit non-variationality. Li$^{+}$ and Be$^{2+}$ were found to have non-variational energies . However, the error is small and within chemical accuracy (within $\sim 0.0016 E_{\text{h}}$).

\section{Graphical analysis of correlation}
\subsection{Electron-electron cusp}
\begin{figure*}[h!]
\centering
\includegraphics[scale=0.7]{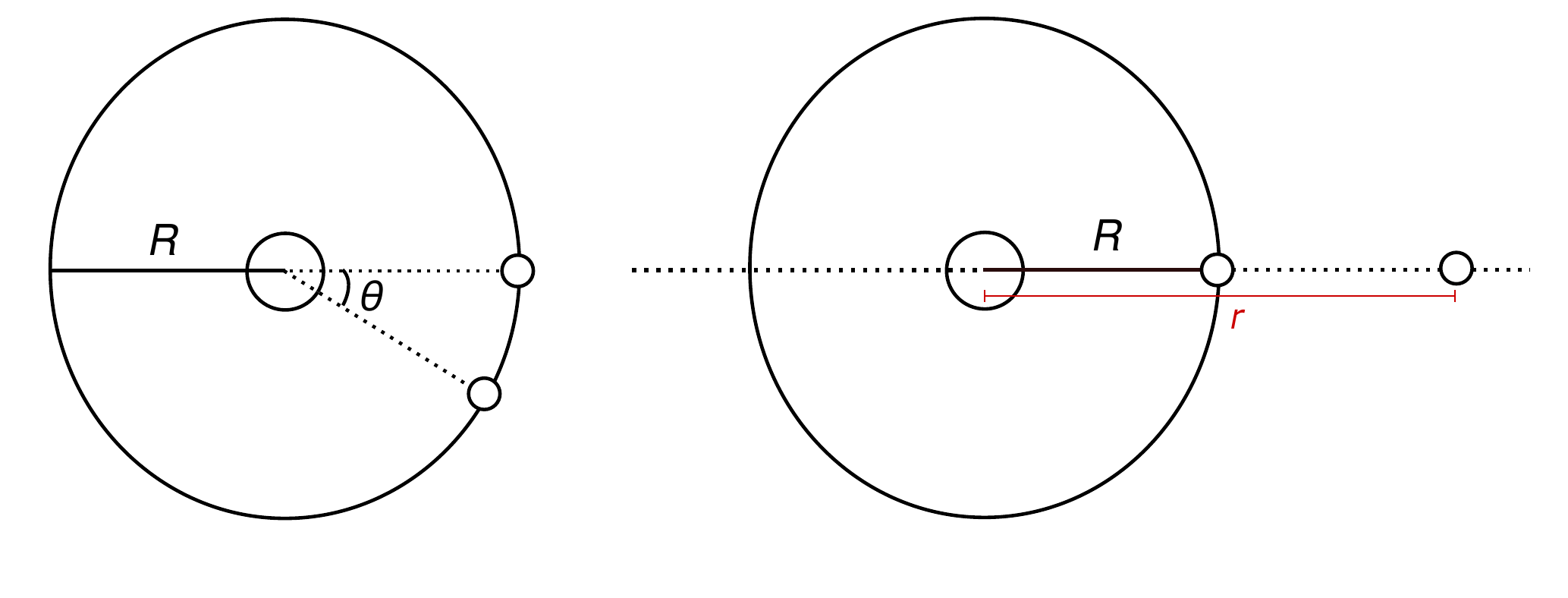}
\caption{(Left) Two electrons constrained at a fixed electron-nuclear distance of $R$. The electron-electron distance is modulated by their angle of separation, $\theta$. $\theta$ is measured in radians. (Right) One electron is fixed at distance $R$ and the other is constrained to the (dotted) line defined by the nucleus and the fixed electron. $R$ is found by the expectation value of the electron-nuclear separation in near-exact wavefunction given by Nakatsuji and coworkers. \autocite{nakashima_solving_2007}}
\label{2e atoms}
\end{figure*}

\begin{figure*}[h!]
\centering
\includegraphics[scale=0.7]{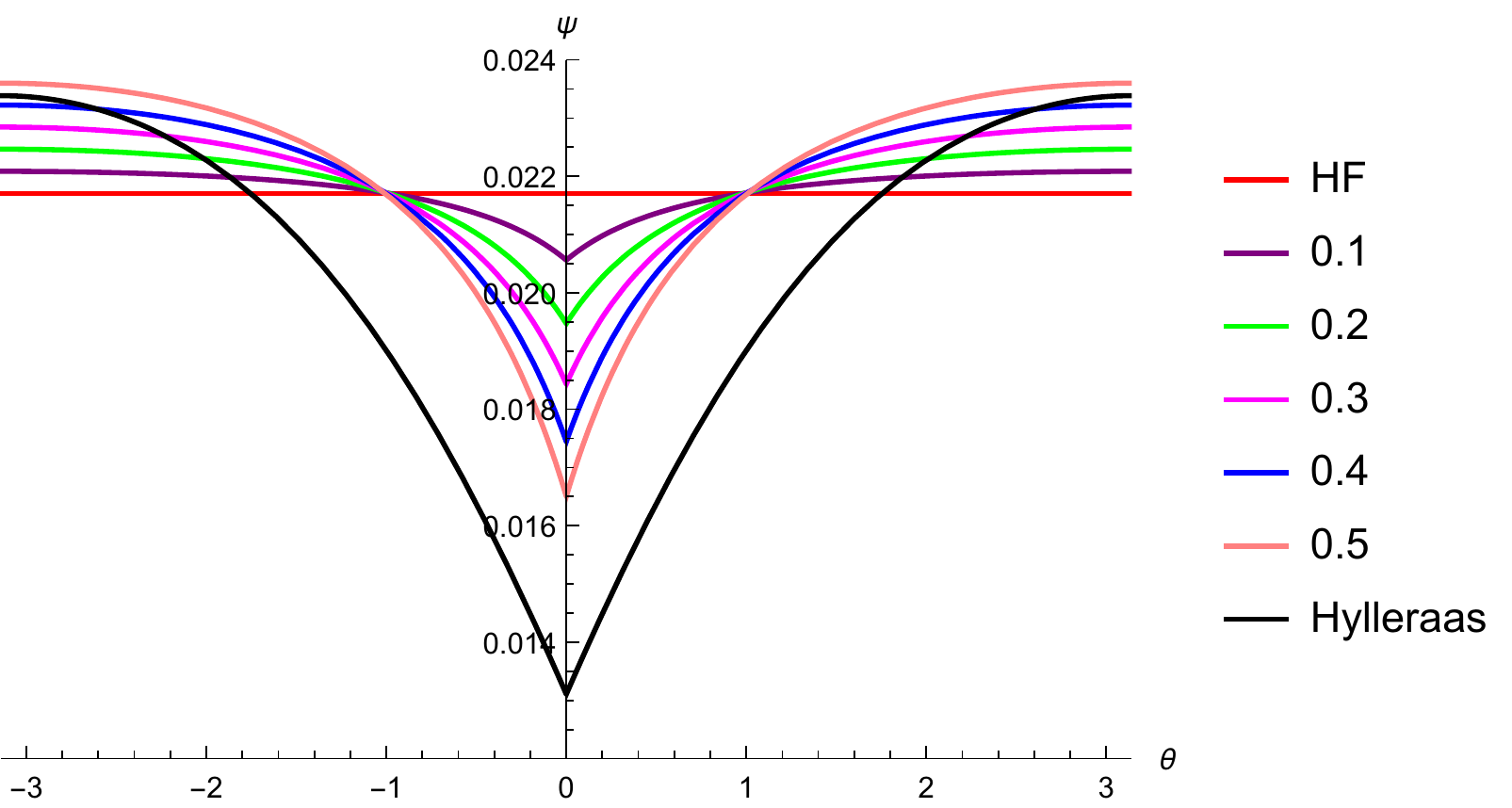}
\caption{Graphical description of the electron-electron cusp of the transcorrelated wavefunction with $c=0.1$ through $c=0.5$ against Hartree--Fock and Hylleraas wavefunctions. The function with $c=0.5$ most closely matches that of the Hylleraas wavefunction, but the cusp is shallow as compared to the Hylleraas wavefunction.}
\label{Angular}
\end{figure*}

To better understand the effects of the electron-electron and electron-nucleus terms in the correlator, we studied the effects of various Jastrow factors $e^{\tau}$ on a Hartree--Fock solution $\Phi_{\text{HF}}$ of a helium atom. We first attempted to study the effects of varying the angle $\theta$ between two electrons confined to the same electron-nucleus distance (Figure \ref{2e atoms}, Left). Using the correlator $\tau = c \frac{r_{12}}{1 + r_{12}}$ and the Slater determinant found with a Hartree--Fock calculation with a cc-pVQZ basis, the transcorrelated wavefunction $e^{\tau} \Phi_{\text{HF}}$ was plotted as a function of $\theta$ (Figure \ref{Angular}) for varying values of the parameter $c$. The 6-term Hylleraas wavefunction \autocite{koga_hylleraas_1990}, which represents a good approximation to the exact wavefunction of helium, is also plotted for comparison.

For ease of reference, the Hylleraas wavefunction for He is given by\autocite{koga_hylleraas_1990}:
\end{multicols}

\begin{center}
\begin{equation}
\Psi_{He} = e^{-1.755656 s} (1 + 0.337294 u + 0.112519 t^{2} - 0.145874 s + 0.023634 s^{2} - 0.037024 u^{2})
\end{equation}
where $s = | \boldsymbol{r}_{1} | +  | \boldsymbol{r}_{2} | $, $t = | \boldsymbol{r}_{1} | -  | \boldsymbol{r}_{2} | $ and $u = | \boldsymbol{r}_{1} - \boldsymbol{r}_{2} | $.\\
\end{center}

\begin{multicols}{2}
The Jastrow factor's introduction of electron-electron cusps to the Hartree--Fock solution can be seen from Figure \ref{Angular}. The shape of the electron-electron cusp gets increasingly similar to that of the Hylleraas wavefunction as the coefficient increases from 0.1 to 0.5. This supports the use of $\tau = \frac{1}{2}\frac{r_{12}}{1 + r_{12}}$ to correct for the electron-electron cusps. The coefficient $c=\frac{1}{2}$ is fixed in the transcorrelated calculations which therefore necessitates the need for higher order terms in $r_{ij}$ to correct for the depth of the cusp.
\end{multicols}

\subsection{Electron-nuclear cusp}

\begin{figure*}[h!]
\centering
\includegraphics[scale=0.7]{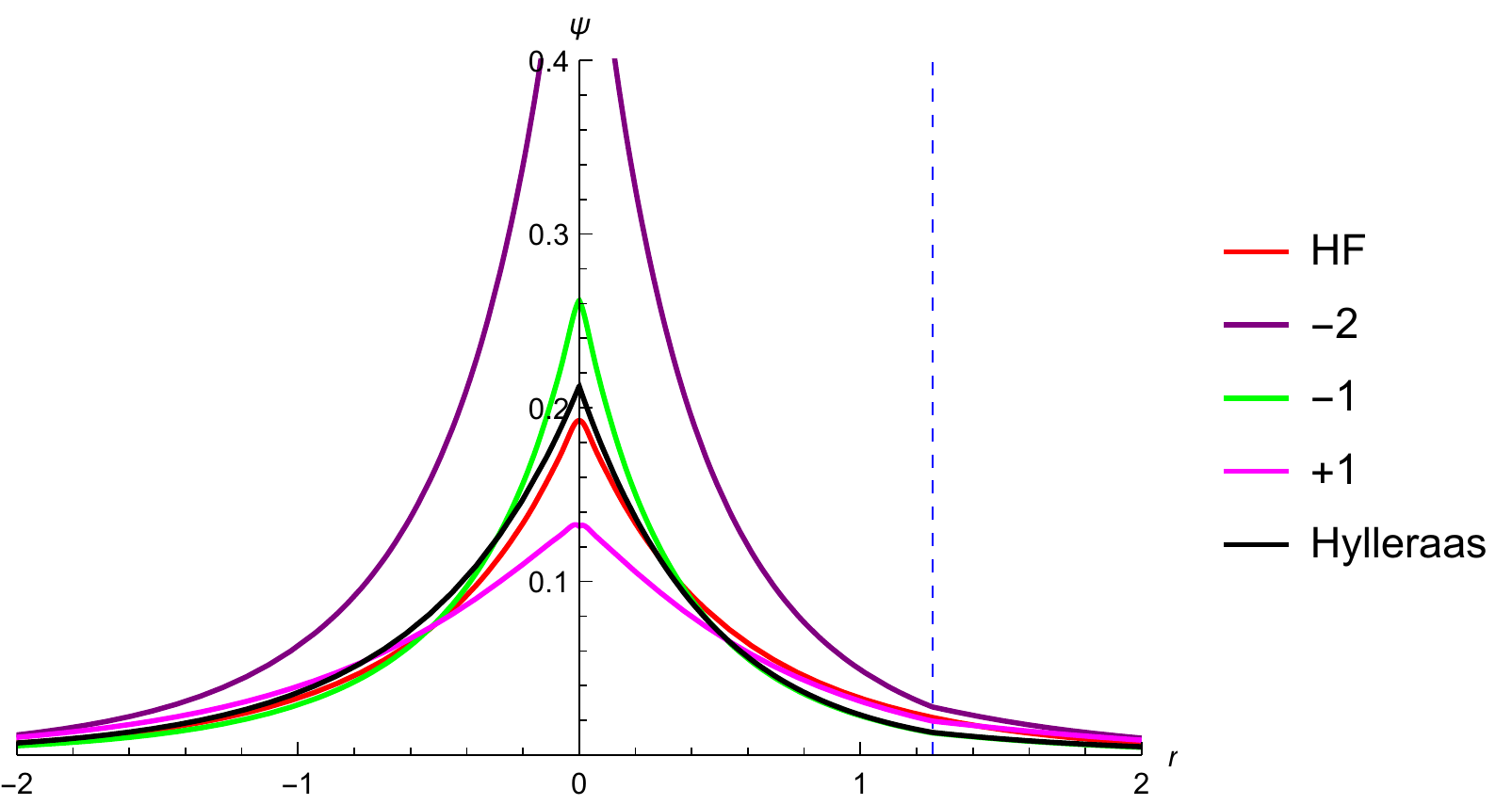}
\caption{Graphical description of the electron-nuclear cusp of the transcorrelated wavefunction of He with $c=-2$, $-1$, and $1$ against Hartree--Fock and Hylleraas wavefunctions. The position at which the other electron is fixed is shown by the dashed blue line.}
\label{He_RadialA}
\end{figure*}

\begin{figure*}[h!]
\centering
\includegraphics[scale=0.7]{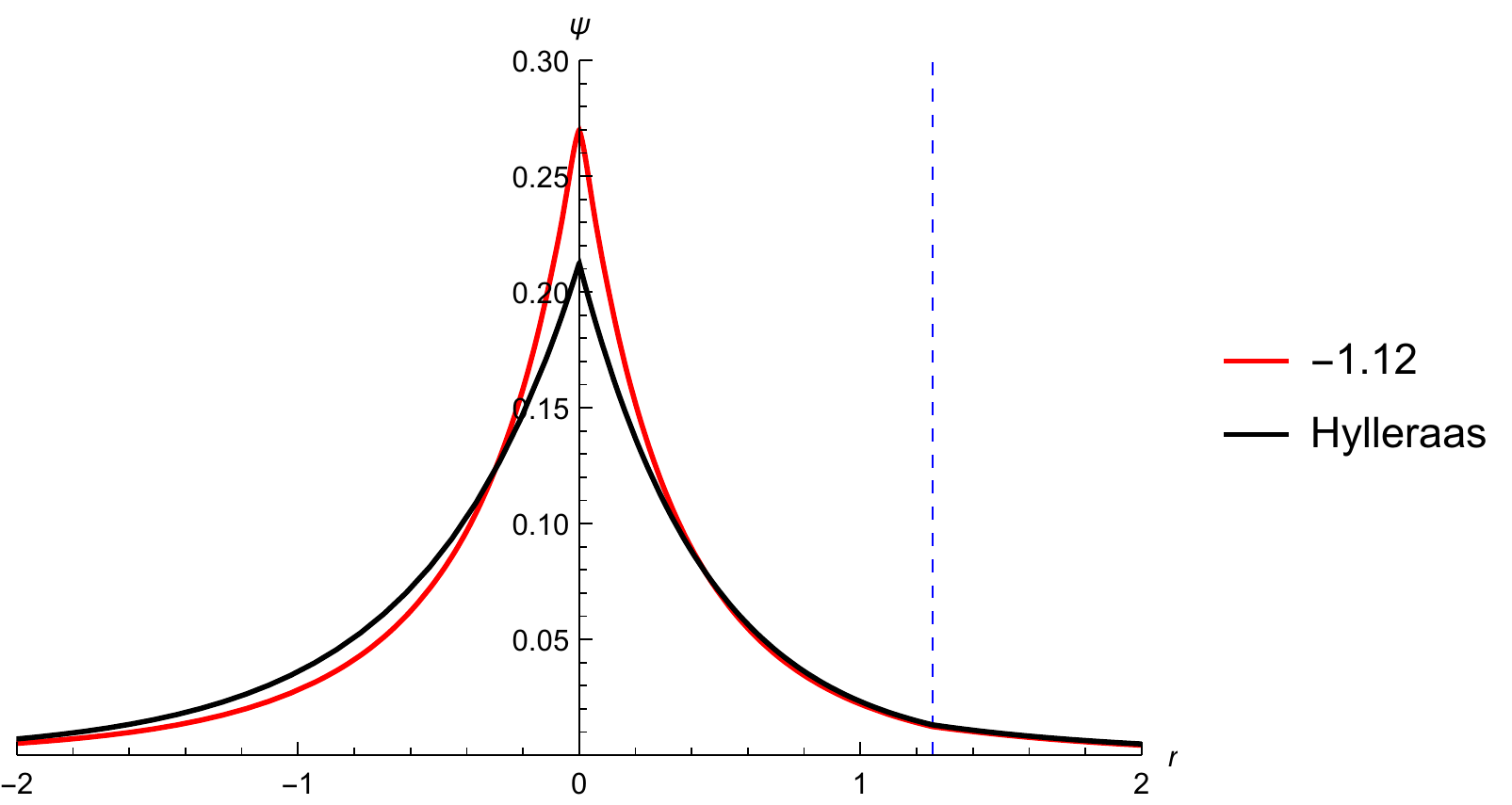}
\caption{Plotting the Hylleraas wavefunction against the transcorrelated wavefunction with $c = -1.12$. $c$ was found by SOM minimisation was performed with a cc-pV5Z basis set. The Slater Determinant used is found via Hartree--Fock calculation using a cc-pVQZ basis. While there is some discrepancy between the transcorrelated wavefunction and the Hylleraas wavefunction near the nucleus, the two wavefunctions are very similar further away from the nucleus. }
\label{He_RadialB}
\end{figure*}

\begin{multicols}{2}

To examine the effect of the electron-nuclear cusp, we use the special case whereby the nucleus and an electron are constrained to be 1.26 Bohr away and the other electron is free to move along the line defined by them (Figure \ref{2e atoms}, Right). We use a different correlator $\tau = \frac{1}{2}\frac{r_{12}}{1 + r_{12}}  + c \frac{r}{1 + r}$ where $r$ is the variable electron-nuclear distance and $r_{12} = r-R$  is the electron-electron distance, and vary the value of parameter $c$. The Hartree--Fock wavefunction most closely matches that of the Hylleraas wavefunction at the nucleus while the function with $c=-1$ is more similar further away from it (Figure \ref{He_RadialA}).

For comparison, the function with $c = -2$ was plotted as $c = -2$ is what would be expected from a simple application of Kato's cusp conditions. This illustrates that the coefficient $c$ need not equal to the negative of the nuclear charge, $-Z$, as the electron-nuclear interaction term in the Jastrow factor affects the overall wavefunction and not only at the cusp. Performing SOM minimisation with this correlator, we found a value of $c = -1.12$, and the transcorrelated wavefunction with $c=-1.12$ is plotted. The plot shows a good agreement with the Hylleraas wavefunction (Figure \ref{He_RadialB}).

A similar series of calculations were attempted for Li$^{+}$. The Hylleraas wavefunction for Li$^{+}$ is given by:
\end{multicols}

\begin{figure*}[h!]
\centering
\includegraphics[scale=0.7]{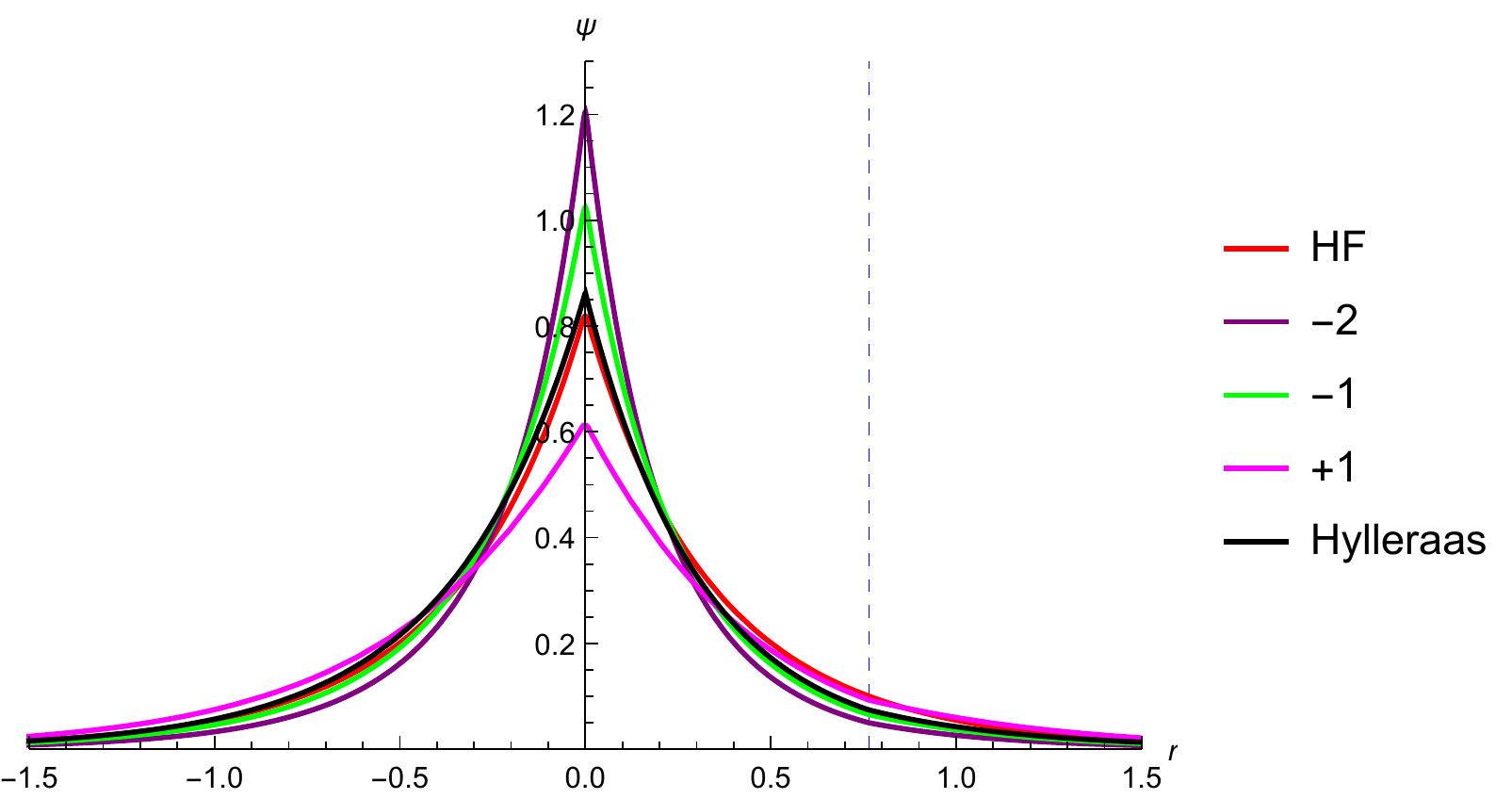}
\caption{Graphical description of the electron-nuclear cusp of the transcorrelated wavefunction of Li$^{+}$ with varying values of $c$ against Hartree--Fock and Hylleraas wavefunctions. The Hartree--Fock wavefunction resembles the Hylleraas wavefunction near the nucleus. The transcorrelated wavefunction with $c=-1$ most closely matches that of the Hylleraas wavefunction further away from the nucleus.}
\label{Li_RadialA}
\end{figure*}

\begin{figure*}[h!]
\centering
\includegraphics[scale=0.7]{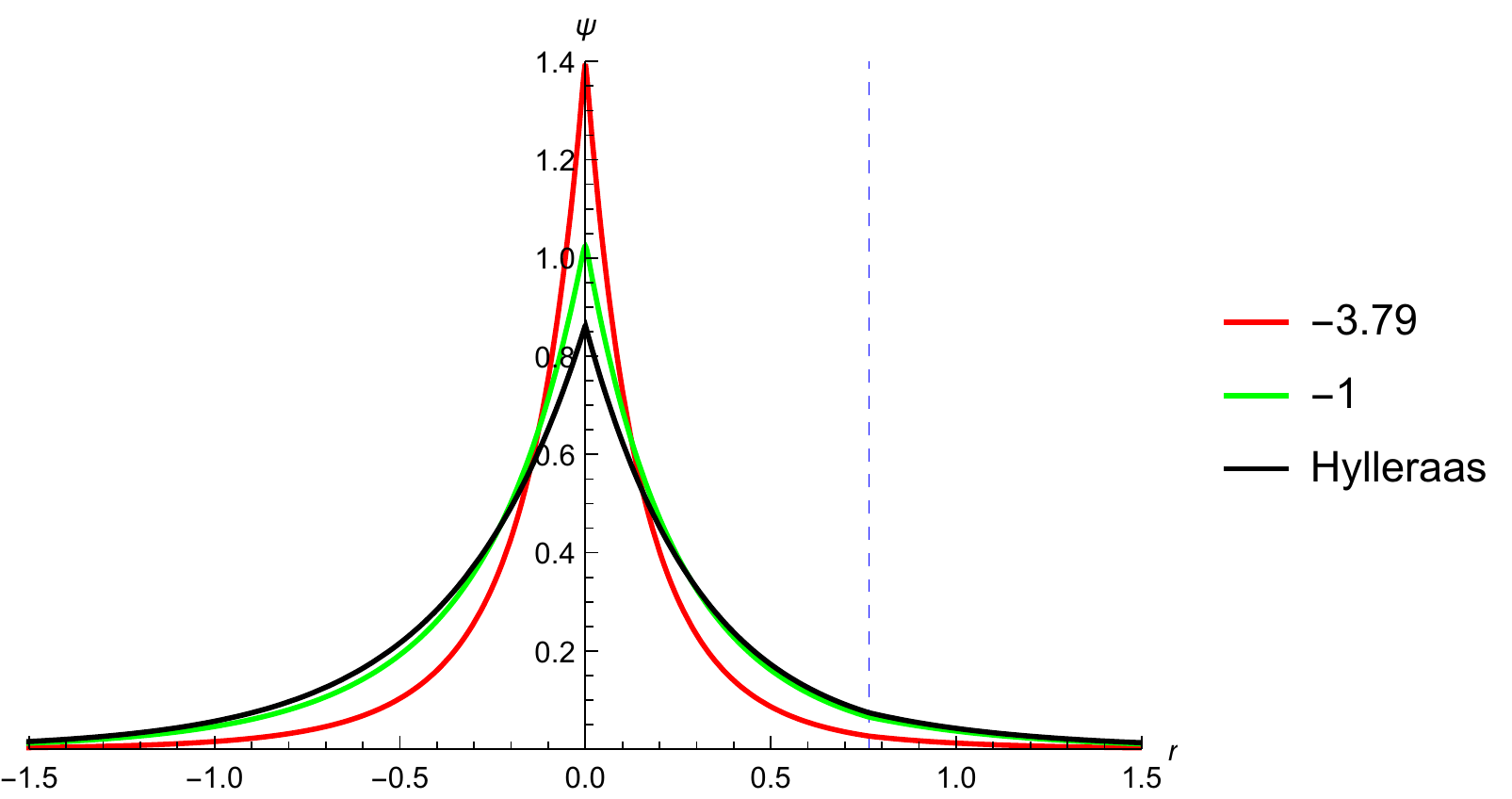}
\caption{Plotting the Hylleraas wavefunction against the transcorrelated wavefunction with $c = -3.79$ and $c=-1$. $c$ was found by SOM minimisation with a cc-pCVQZ basis set. The Slater Determinant used is found via Hartree--Fock calculation using a cc-pVQZ basis.}
\label{Li_RadialB}
\end{figure*}

\begin{figure*}[h!]
\centering
\includegraphics[scale=0.5]{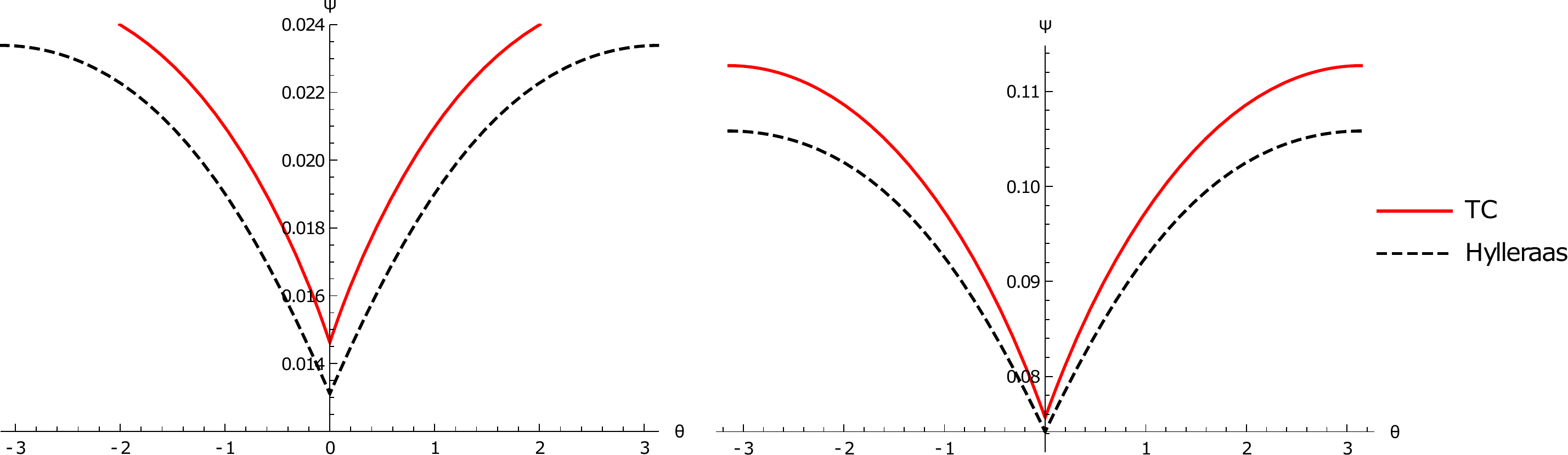}
\caption{Two plots describing the electron-electron cusp corresponding to Figure \ref{2e atoms} (Left).\\
(Left) Plot of transcorrelated wavefunction for He against the corresponding Hylleraas wavefunction. (Right) Plot of transcorrelated wavefunction for Li$^{+}$ against the corresponding Hylleraas wavefunction. Both plots show that the 18 parameter transcorrelated wavefunction reproduces the shape of the electron-electron cusp, but the cusp is shallower than the Hylleraas wavefunction.}
\label{TCAngular}
\end{figure*}

\begin{figure*}[h!]
\centering
\includegraphics[scale=0.5]{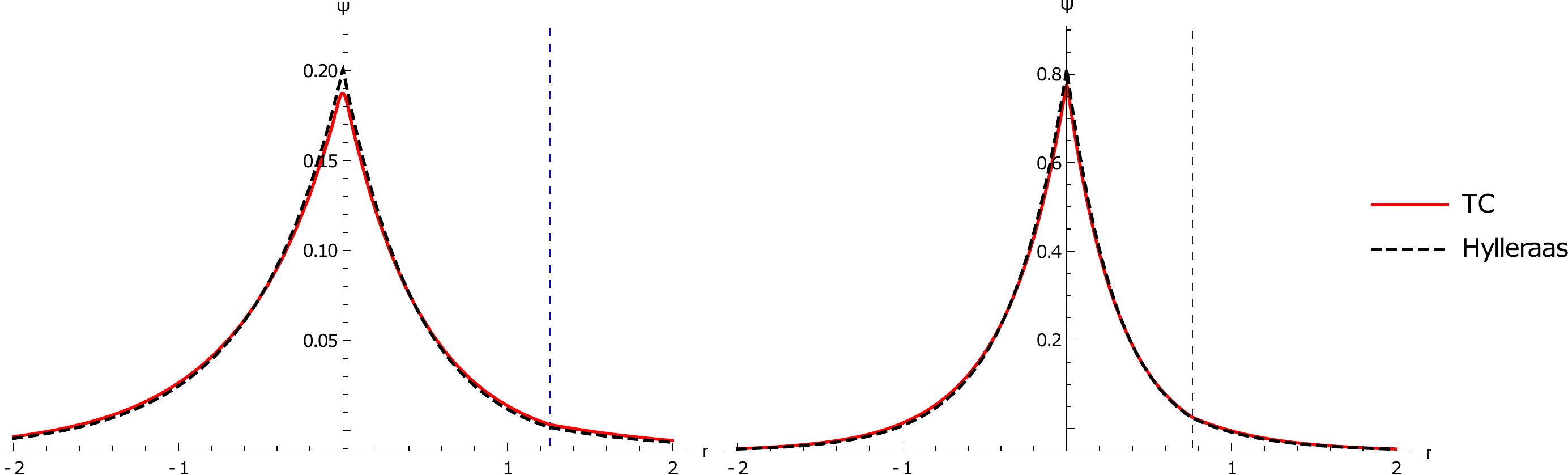}
\caption{Two plots describing the electron-nuclear cusp corresponding to Figure \ref{2e atoms} (Right). \\
(Left) Plot of transcorrelated wavefunction for He against the corresponding Hylleraas wavefunction. (Right) Plot of transcorrelated wavefunction for Li$^{+}$ against the corresponding Hylleraas wavefunction. Both plots show that the 18 parameter transcorrelated wavefunction reproduces the Hylleraas wavefunction well.}
\label{TCRadial}
\end{figure*}

\begin{equation}
\Psi_{\text{Li}^{+}} = e^{-2.784751 s} (1 + 0.354317 u + 0.154657 t^{2} - 0.127225 s + 0.042220 s^{2} - 0.066731 u^{2})
\end{equation}

\begin{multicols}{2}
From Figure \ref{Li_RadialA} it can be seen that the Hartree--Fock wavefunction is very similar to that of the Hylleraas wavefunction near the nucleus. At regions further from the nucleus, the $c= -1$ wavefunction most closely resembles the Hylleraas wavefunction. SOM minimisation gave $c = -3.79$ but in this case, we were unable to reproduce the Hylleraas wavefunction (Figure \ref{Li_RadialB}). 
There are several reasons why such a discrepancy can exist.

Firstly, resolution of identity is an \emph{approximation} which may not be valid depending on the size of the basis set used.

Secondly, the Hartree--Fock wavefunction resembles the Hylleraas wavefunction well. While the addition of a Jastrow factor to it can provide a better description of the electron-nuclear cusp, it comes at the cost of affecting other parts of the wavefunction. More terms in the correlator may need to be added to more accurately describe the transcorrelated at different points in space.

To address the latter point, the transcorrelated wavefunctions for He and Li$^{+}$ found using the SOM18 set of parameters (Table \ref{Results: HF-SOM}) were plotted against the respective Hylleraas wavefunctions (Figures \ref{TCAngular}, \ref{TCRadial}). From Table \ref{Results: HF-SOM}, it can be observed that the transcorrelated energies are very close to the exact energies, and this suggests that the transcorrelated wavefunctions should look similar to that of the Hylleraas wavefunction. This is reflected graphically in the depiction of electron-electron cusp (Figure \ref{TCAngular}) and electron-nuclear cusp (Figure \ref{TCRadial}), where each of the transcorrelated wavefunctions agree well with the corresponding Hylleraas wavefunction. The main difference between the transcorrelated wavefunction and Hylleraas wavefunctions appear to be the description of electron-electron interactions at larger electron-electron distances, which may hint at the use of a differently scaled form of the correlator to account for longer range effects. Overall, the plots shows that SOM minimisation can get highly accurate wavefunctions given sufficiently many correlator parameters, supporting the utility of SOM minimisation when appropriate starting guesses are used.

\section{Conclusions}
A self-consistent method for solving the non-self-adjoint transcorrelated Hamiltonian has been implemented successfully to obtain highly accurate energies of some first row atoms. The correlator parameters found in the literature are not optimised for the Gaussian orbital basis used in this current study and had to be re-optimised through a method we refer to as SOM minimisation. This allowed us to find optimised parameters for any system, in principle. However, the optimisation of multiple parameters is challenging and in practice, we have found it to be useful to optimise the parameters using a two-step SOM minimisation procedure. SOM minimisation has been found to give good energies for the first row atoms. However, the percentage of correlation energy recovered has been found to decrease with increased nuclear charge across a series of helium-like ions. We believe that this is due to the inability of the basis set to accurately describe highly charged cations and a custom basis with more contracted basis functions would be helpful to describe the correlation in these systems.\\
\newline
Thus far, SOM minimisation has been attempted for closed shell systems. Further work has to be done on open-shell systems where there is a possibility of spin-symmetry breaking, leading to an unphysical wavefunction. SOM minimisation should also be attempted on larger systems to test if the method works more generally. However, as pointed out by Alavi and coworkers\autocite{cohen_similarity_2019}, memory use is a bottleneck in transcorrelated calculations. This can be challenging especially with the need to use large basis sets for SOM minimisation as it relies on the resolution-of-the-identity approximation. A possible solution would be to use an auxiliary basis set instead which would allow us to use a smaller basis set to represent the Slater Determinant but a large auxiliary basis to satisfy the resolution-of-the-identity approximation. A graphical analysis has also been done to illustrate the effects of some correlator terms on the overall wavefunction and demonstrated the importance of including higher-order correlator terms in the Jastrow factor to give a more accurate wavefunction.\\
\newline

\section{Acknowledgements}
NL is grateful for the helpful feedback and insights provided by Prof. Ali Alavi and Prof. Matthew Foulkes.\\
\clearpage

\end{multicols}

\appendix

\begin{multicols}{2}

\section{Transcorrelated Hamiltonian}
\label{Appendix A}
\subsection{Expansion of commutator terms}
We first separate the many-electron electronic Hamiltonian into a kinetic energy ($\hat{T}$) and potential energy ($\hat{V}$) term:
\begin{equation}
\hat{H} = \underbrace{ - \frac{1}{2} \sum_{i} \nabla_{i}^{2} }_\text{$\hat{T}$}- \underbrace{\sum_{iA} \frac{Z_{A}}{ |\boldsymbol{r}_{1} - \boldsymbol{R}_{A} |} + \sum_{i<j}\frac{1}{| \boldsymbol{r}_{i} - \boldsymbol{r}_{j} | } }_\text{$\hat{V}$}
\end{equation}
$\hat{V}$ is multiplicative and hence commutes with $\tau$ in the commutator $[\hat{H}, \tau]$, that is,
\begin{equation}
\begin{split}
	[\hat{H}, \tau] &= [\hat{T} + \hat{V}, \tau] \\
						   &= [\hat{T}, \tau]
\end{split}
\end{equation}
The commutator terms in equation \ref{TC: BCH expansion} will be evaluated term by term as follows:
\begin{equation}
\begin{split}
	[\hat{H}, \tau]f &= [\hat{T}, \tau]f \\
						    &= \hat{T}(\tau f) - \tau \hat{T} f \\
						    &= -\frac{1}{2} \sum_{i} \boldsymbol{\nabla}_{i} \cdot \boldsymbol{\nabla}_{i} (\tau f) + \frac{1}{2} \sum_{i} \tau \nabla_{i}^{2} f \\
						    &= -\frac{1}{2} \sum_{i} \boldsymbol{\nabla}_{i} \cdot (\tau \boldsymbol{\nabla}_{i} f + f \boldsymbol{\nabla}_{i} \tau) + \frac{1}{2} \sum_{i} \tau \nabla_{i}^{2} f \\
						    &= -\frac{1}{2} \sum_{i} (\tau \nabla_{i}^{2} f + f \nabla_{i}^{2} \tau + 2 \boldsymbol{\nabla}_{i} \tau \cdot \boldsymbol{\nabla}_{i} f ) \\
						    &\quad + \frac{1}{2} \sum_{i} \tau \nabla_{i}^{2} f \\
						    &= \Big( -\frac{1}{2} \sum_{i} \nabla_{i}^{2} \tau - \sum_{i} \boldsymbol{\nabla}_{i} \tau \cdot \boldsymbol{\nabla}_{i} \Big) f\\
\end{split}
\end{equation}
We can now make the identification
\begin{equation}
	[\hat{H}, \tau] \equiv -\sum_{i} \Big(\frac{1}{2} \nabla_{i}^{2} \tau + \boldsymbol{\nabla}_{i} \tau \cdot \boldsymbol{\nabla}_{i} \Big)
\end{equation}

\begin{equation}
\begin{split}
	[[\hat{H}, \tau], \tau]f &= [\hat{H}, \tau](\tau f) - \tau [\hat{H}, \tau] f \\
									   &= \sum_{i} \Big( - \boldsymbol{\nabla}_{i} \tau \cdot \boldsymbol{\nabla}_{i} (\tau f) + \tau \boldsymbol{\nabla}_{i}\boldsymbol \tau \cdot \boldsymbol{\nabla}_{i} f \Big) \\
									   &= \sum_{i} \Big( - f \boldsymbol{\nabla}_{i} \tau \cdot \boldsymbol{\nabla}_{i} \tau - \tau \boldsymbol{\nabla}_{i} \tau \cdot \boldsymbol{\nabla}_{i} f \\
									   &\quad + \tau \boldsymbol{\nabla}_{i} \tau \cdot \boldsymbol{\nabla}_{i} f \Big) \\
									   &= \sum_{i} \Big( - f \boldsymbol{\nabla}_{i} \tau \cdot \boldsymbol{\nabla}_{i} \tau \Big) \\
									   &= \sum_{i} - \Big(\boldsymbol{\nabla}_{i} \tau \Big)^{2} f \\ 
\end{split}
\end{equation}
Therefore,
\begin{equation}
	[[\hat{H}, \tau], \tau] \equiv \sum_{i} - \Big(\boldsymbol{\nabla}_{i} \tau \Big)^{2}
\end{equation}
Since $[[\hat{H}, \tau], \tau]$ is a multiplicative term, higher-order commutators of the form $[[[\hat{H}, \tau], \tau] ... ]$ vanish.

Given that $\tau = \sum_{i<j} u (\boldsymbol{r}_{i}, \boldsymbol{r}_{j})$, we can further expand equation \ref{TC: TC expanded}. \\
\begin{equation}
\begin{split}
	\boldsymbol{\nabla}_{i} \tau &= \boldsymbol{\nabla}_{i} \sum_{a<b} u (\boldsymbol{r}_{a}, \boldsymbol{r}_{b}) \\
												  &= \sum_{a} \boldsymbol{\nabla}_{i} u (\boldsymbol{r}_{i}, \boldsymbol{r}_{a}) \\
\end{split}
\end{equation}
where we used the symmetry $u (\boldsymbol{r}_{i}, \boldsymbol{r}_{j}) = u (\boldsymbol{r}_{j}, \boldsymbol{r}_{i})$.
We therefore find:
\begin{equation}
\begin{split}
	\sum_{i} \nabla_{i}^{2} \tau 	&=  \sum_{i}\sum_{a} \nabla_{i}^{2} u (\boldsymbol{r}_{i}, \boldsymbol{r}_{a}) \\
													&= \frac{1}{2}\sum_{i}\sum_{a} \nabla_{i}^{2} u (\boldsymbol{r}_{i}, \boldsymbol{r}_{a}) \\
													&\quad + \frac{1}{2}\sum_{i}\sum_{a} \nabla_{a}^{2} u (\boldsymbol{r}_{a}, \boldsymbol{r}_{i})\\
													&= \sum_{i<j} \nabla_{i}^{2} u (\boldsymbol{r}_{i}, \boldsymbol{r}_{j}) + \sum_{i<j} \nabla_{j}^{2} u (\boldsymbol{r}_{i}, \boldsymbol{r}_{j})
\end{split}
\end{equation}
where we relabelled $a$ by $j$ and used the symmetry of $u$ in the last line.\\
We similarly find:
\begin{equation}
\begin{split}
	\sum_{i}\boldsymbol{\nabla}_{i} \tau \cdot \boldsymbol{\nabla}_{i} &=\sum_{i}\sum_{a} \boldsymbol{\nabla}_{i} u (\boldsymbol{r}_{i}, \boldsymbol{r}_{a}) \cdot \boldsymbol{\nabla}_{i} \\
																											   &= \frac{1}{2}\sum_{i}\sum_{a} \boldsymbol{\nabla}_{i} u (\boldsymbol{r}_{i}, \boldsymbol{r}_{a}) \cdot \boldsymbol{\nabla}_{i} \\
																											   &\quad + \frac{1}{2}\sum_{i}\sum_{a} \boldsymbol{\nabla}_{a} u (\boldsymbol{r}_{a}, \boldsymbol{r}_{i}) \cdot \boldsymbol{\nabla}_{a} \\
																											   &= \sum_{i<j} \boldsymbol{\nabla}_{i} u (\boldsymbol{r}_{i}, \boldsymbol{r}_{j}) \cdot \boldsymbol{\nabla}_{i} + \sum_{i<j} \boldsymbol{\nabla}_{j} u (\boldsymbol{r}_{i}, \boldsymbol{r}_{j}) \cdot \boldsymbol{\nabla}_{j} 
\end{split}
\end{equation}
Finally,
\begin{equation}
\begin{split}
	\sum_{i} \Big(\boldsymbol{\nabla}_{i} \tau \Big)^{2} &= \sum_{i} \boldsymbol{\nabla}_{i} \tau \cdot \boldsymbol{\nabla}_{i} \tau \\
																					   &= \sum_{i} \sum_{a} \boldsymbol{\nabla}_{i} u (\boldsymbol{r}_{i}, \boldsymbol{r}_{a}) \cdot  \sum_{b} \boldsymbol{\nabla}_{i} u (\boldsymbol{r}_{i}, \boldsymbol{r}_{b})  \\
																					   &= \sum_{i, j} \boldsymbol{\nabla}_{i} u (\boldsymbol{r}_{i}, \boldsymbol{r}_{j}) \cdot \boldsymbol{\nabla}_{i} u (\boldsymbol{r}_{i}, \boldsymbol{r}_{j}) \\
																					   &\quad + \sum_{i, a \neq b} \boldsymbol{\nabla}_{i} u (\boldsymbol{r}_{i}, \boldsymbol{r}_{a}) \cdot \boldsymbol{\nabla}_{i} u (\boldsymbol{r}_{i}, \boldsymbol{r}_{b})\\
																					   &= \sum_{i<j} \Big(\boldsymbol{\nabla}_{i} u (\boldsymbol{r}_{i}, \boldsymbol{r}_{j}) \Big)^{2} + \sum_{i<j} \Big(\boldsymbol{\nabla}_{j} u (\boldsymbol{r}_{i}, \boldsymbol{r}_{j})  \Big)^{2} \\
																					   &\quad+ \sum_{i}\sum_{a<b} \Big(\boldsymbol{\nabla}_{i} u (\boldsymbol{r}_{i}, \boldsymbol{r}_{a}) \cdot \boldsymbol{\nabla}_{i} u (\boldsymbol{r}_{i}, \boldsymbol{r}_{b})\Big) \\
																					   &\quad+ \sum_{i}\sum_{b<a} \Big(\boldsymbol{\nabla}_{i} u (\boldsymbol{r}_{i}, \boldsymbol{r}_{a}) \cdot \boldsymbol{\nabla}_{i} u (\boldsymbol{r}_{i}, \boldsymbol{r}_{b})\Big) \\
																					   &= \sum_{i<j} \Big(\boldsymbol{\nabla}_{i} u (\boldsymbol{r}_{i}, \boldsymbol{r}_{j}) \Big)^{2} + \sum_{i<j} \Big(\boldsymbol{\nabla}_{j} u (\boldsymbol{r}_{i}, \boldsymbol{r}_{j})  \Big)^{2} \\
																					   &\quad+ 2 \sum_{i}\sum_{a<b} \Big(\boldsymbol{\nabla}_{i} u (\boldsymbol{r}_{i}, \boldsymbol{r}_{a}) \cdot \boldsymbol{\nabla}_{i} u (\boldsymbol{r}_{i}, \boldsymbol{r}_{b})\Big) \\
																					   &= \sum_{i<j} \Big(\boldsymbol{\nabla}_{i} u (\boldsymbol{r}_{i}, \boldsymbol{r}_{j}) \Big)^{2} + \sum_{i<j} \Big(\boldsymbol{\nabla}_{j} u (\boldsymbol{r}_{i}, \boldsymbol{r}_{j})  \Big)^{2} \\
																					   &\quad+ 2\sum_{i<a<b} \Big(\boldsymbol{\nabla}_{i} u (\boldsymbol{r}_{i}, \boldsymbol{r}_{a}) \cdot \boldsymbol{\nabla}_{i} u (\boldsymbol{r}_{i}, \boldsymbol{r}_{b})\Big) \\
																					   &\quad+ 2\sum_{a<i<b} \Big(\boldsymbol{\nabla}_{i} u (\boldsymbol{r}_{i}, \boldsymbol{r}_{a}) \cdot \boldsymbol{\nabla}_{i} u (\boldsymbol{r}_{i}, \boldsymbol{r}_{b})\Big) \\
																					    &\quad+ 2\sum_{a<b<i} \Big(\boldsymbol{\nabla}_{i} u (\boldsymbol{r}_{i}, \boldsymbol{r}_{a}) \cdot \boldsymbol{\nabla}_{i} u (\boldsymbol{r}_{i}, \boldsymbol{r}_{b})\Big) \\
																					    &= \sum_{i<j} \Big(\boldsymbol{\nabla}_{i} u (\boldsymbol{r}_{i}, \boldsymbol{r}_{j}) \Big)^{2} + \sum_{i<j} \Big(\boldsymbol{\nabla}_{j} u (\boldsymbol{r}_{i}, \boldsymbol{r}_{j})  \Big)^{2} \\
																					   &\quad+ 2\sum_{i<j<k} \Big(\boldsymbol{\nabla}_{i} u (\boldsymbol{r}_{i}, \boldsymbol{r}_{j}) \cdot \boldsymbol{\nabla}_{i} u (\boldsymbol{r}_{i}, \boldsymbol{r}_{k}) \\
																					   &\quad + \boldsymbol{\nabla}_{j} u (\boldsymbol{r}_{j}, \boldsymbol{r}_{i}) \cdot \boldsymbol{\nabla}_{j} u (\boldsymbol{r}_{j}, \boldsymbol{r}_{k}) \\
																					   	&\quad\quad\quad\quad\quad+\boldsymbol{\nabla}_{k} u (\boldsymbol{r}_{k}, \boldsymbol{r}_{i}) \cdot \boldsymbol{\nabla}_{k} u (\boldsymbol{r}_{k}, \boldsymbol{r}_{j}) \Big)
\end{split}
\end{equation}
Substituting these commutator terms back into equation \ref{TC: BCH expansion} recovers equation \ref{TC: TC expanded}.
\subsection{Lagrangian approach to the Transcorrelated Hamiltonian}
We show here that the method of Lagrange multipliers can be used to derive the effective transcorrelated Hamiltonian.
\begin{equation}
\begin{split}
\mathcal{L} &= E - \sum_{i=1}\sum_{j=1} \epsilon_{ij} (\braket{ \psi_{i} | \phi_{j} } -\delta_{ij} )\\
				   &= \braket{ \Psi | \bar{H} | \Phi } - \sum_{i=1}\sum_{j=1} \epsilon_{ij} (\braket{ \psi_{i} | \phi_{j} } -\delta_{ij} )\\
				   &= \braket{ \Psi | \hat{H} - \sum_{i<j} \hat{K}(\boldsymbol{r}_{i}, \boldsymbol{r}_{j}) - \sum_{i<j<k} \hat{L}(\boldsymbol{r}_{i}, \boldsymbol{r}_{j}, \boldsymbol{r}_{k}) | \Phi } \\
				   &\quad - \sum_{i=1}\sum_{j=1} \epsilon_{ij} (\braket{ \psi_{i} | \phi_{j} } -\delta_{ij} )\\
				   &= \braket{ \Psi | \sum_{i} \hat{h}_{i} | \Phi } + \braket{ \Psi | \sum_{i<j} ( r_{ij}^{-1} -  \hat{K}(\boldsymbol{r}_{i}, \boldsymbol{r}_{j}) ) | \Phi } \\
				   &\quad - \braket{ \Psi | \sum_{i<j<k} \hat{L}(\boldsymbol{r}_{i}, \boldsymbol{r}_{j}, \boldsymbol{r}_{k}) | \Phi }  \\
				   &\quad- \sum_{i=1}\sum_{j=1} \epsilon_{ij} (\braket{ \psi_{i} | \phi_{j} } -\delta_{ij} )\\
				   &= \braket{ \Psi | \hat{O}_{1} | \Phi } + \braket{ \Psi | \hat{O}_{2} | \Phi } + \braket{ \Psi | \hat{O}_{3} | \Phi }  \\
				   &\quad - \sum_{i=1}\sum_{j=1} \epsilon_{ij} (\braket{ \psi_{i} | \phi_{j} } -\delta_{ij} ) \\
\end{split}
\end{equation}
In the last line we have renamed the $n$-electron operators by the $\hat{O}_{n}$. This is for brevity of notation and the understanding that the mathematics after is concerned only with the number of electrons the operators act upon. Taking an infinitesimal change of the Lagrangian with respect to variation of the spin-orbitals,
\begin{equation}
\begin{split}
\delta \mathcal{L} &= \delta \braket{ \Psi | \hat{O}_{1} | \Phi } + \delta \braket{ \Psi | \hat{O}_{2} | \Phi } + \delta \braket{ \Psi | \hat{O}_{3} | \Phi }  \\
							  &\quad - \delta \sum_{i=1}\sum_{j=1} \epsilon_{ij} (\braket{ \psi_{i} | \phi_{j} } -\delta_{ij} )
\end{split}
\end{equation}
We shall analyse each term of the above expression in turn. We will adopt the shorthand $\braket{ \psi_{i} \psi_{j} ... \psi_{k} | \hat{O} | \phi_{a} \phi_{b} ... \phi_{c} } = \braket{ ij ... k | \hat{O} | ab ... c}$. We assume that the bra always contains molecular orbitals from the set $\{\psi_{i} \}$ and that the ket always contains molecular orbitals from the set $\{\phi_{i} \}$.
\subsubsection{One electron term}
\begin{equation}
\begin{split}
	\delta \braket{ \Psi | \hat{O}_{1} | \Phi } &= \delta \Big( \sum_{i} \braket{ i | \hat{O}_{1} | i } \Big) \\
																	&= \sum_{i} \Big( \braket{ \delta i | \hat{O}_{1} | i } + \braket{ i | \hat{O}_{1} | \delta i } \Big)
\end{split}
\end{equation}
\subsubsection{Two electron term}
\begin{equation}
\begin{split}
	\delta \braket{ \Psi | \hat{O}_{2} | \Phi } &= \sum_{i<j} \Big( \delta \braket{ i j | \hat{O}_{2} | i j } - \delta \braket{ i j | \hat{O}_{2} | j i } \Big)\\
																	&= \sum_{i<j} \Big( \braket{ \delta i j | \hat{O}_{2} | i j } + \braket{ i \delta j | \hat{O}_{2} | i j } \\
																	&\quad + \braket{ i j | \hat{O}_{2} | \delta i j } + \braket{ i j | \hat{O}_{2} | i \delta j } \\
																	&\quad - \braket{ \delta i j | \hat{O}_{2} | j i } - \braket{ i \delta j | \hat{O}_{2} | j i } \\
																	&\quad - \braket{ i j | \hat{O}_{2} | \delta j i } - \braket{ i j | \hat{O}_{2} | j \delta i } \Big)\\
																	&= \sum_{i<j} \Big( 2 \braket{ \delta i j | \hat{O}_{2} | i j } - 2 \braket{ \delta i j | \hat{O}_{2} | j i } \\
																	&\quad + 2 \braket{ i j | \hat{O}_{2} | \delta i j } - 2 \braket{ j i | \hat{O}_{2} | \delta i j } \Big)\\
\end{split}
\end{equation}
Where we have used the permutation symmetry of the integrals e.g. $\braket{ i j | \hat{O}_{2} | i j } =  \braket{ j i | \hat{O}_{2} | j i }$.

\subsubsection{Three electron term}
\begin{equation}
\begin{split}
	\delta \braket{ \Psi | \hat{O}_{3} | \Phi } &= \sum_{i<j<k} \Big( \delta \braket{ i j k | \hat{O}_{3} | i j k } - \delta \braket{ i j k | \hat{O}_{3} | i k j } \\
																	&\quad + \delta \braket{ i j k | \hat{O}_{3} | j k i } - \delta \braket{ i j k | \hat{O}_{3} | j i k } \\
																	&\quad + \delta \braket{ i j k | \hat{O}_{3} | k i j } - \delta \braket{ i j k | \hat{O}_{3} | k j i } \Big)\\
																	&= \sum_{i<j<k} \Big( 3 \braket{ \delta i j k | \hat{O}_{3} | i j k } - 3 \braket{ \delta i j k | \hat{O}_{3} | i k j } \\
																	&\quad + 3 \braket{ \delta i j k | \hat{O}_{3} | j k i } - 3 \braket{ \delta i j k | \hat{O}_{3} | j i k } \\
																	&\quad + 3 \braket{ \delta i j k | \hat{O}_{3} | k i j } - 3 \braket{ \delta i j k | \hat{O}_{3} | k j i } \\
																	&\quad + 3 \braket{ i j k | \hat{O}_{3} | \delta i j k } - 3 \braket{ i k j | \hat{O}_{3} | \delta i j k } \\
																	&\quad + 3 \braket{ j k i | \hat{O}_{3} | \delta i j k } - 3 \braket{ j i k | \hat{O}_{3} | \delta i j k } \\
																	&\quad + 3 \braket{ k i j | \hat{O}_{3} | \delta i j k } - 3 \braket{ k j i | \hat{O}_{3} | \delta i j k } \Big)\\
\end{split}
\end{equation}

\subsubsection{Lagrangian differential}
\begin{equation}
\begin{split}
	\delta \mathcal{L} 	&= \delta \braket{ \Psi | \hat{O}_{1} | \Phi } + \delta \braket{ \Psi | \hat{O}_{2} | \Phi } \\
									&\quad+ \delta \braket{ \Psi | \hat{O}_{3} | \Phi }  - \delta \sum_{i=1}\sum_{j=1} \epsilon_{ij} (\braket{ i | j } -\delta_{ij} ) \\
								  	&= \delta \braket{ \Psi | \hat{O}_{1} | \Phi } + \delta \braket{ \Psi | \hat{O}_{2} | \Phi } \\
								  	&\quad + \delta \braket{ \Psi | \hat{O}_{3} | \Phi }  - \sum_{i=1}\sum_{j=1} \epsilon_{ij} (\braket{ \delta i | j } + \braket{ i | \delta j } ) \\
								  	&= \delta \mathcal{L}_{\psi} + \delta \mathcal{L}_{\phi}
\end{split}
\end{equation}
where we have defined the following:
\begin{equation}
\begin{split}
	\delta \mathcal{L}_{\psi} &= \sum_{i} \braket{ \delta i | \hat{O}_{1} | i } \\
											&\quad + \sum_{i<j} \Big( 2 \braket{ \delta i j | \hat{O}_{2} | i j } - 2 \braket{ \delta i j | \hat{O}_{2} | j i } \Big) \\
											&\quad + \sum_{i<j<k} \Big( 3 \braket{ \delta i j k | \hat{O}_{3} | i j k } - 3 \braket{ \delta i j k | \hat{O}_{3} | i k j } \\
											&\quad	 + 3 \braket{ \delta i j k | \hat{O}_{3} | j k i } - 3 \braket{ \delta i j k | \hat{O}_{3} | j i k } \\
											&\quad + 3 \braket{ \delta i j k | \hat{O}_{3} | k i j } - 3 \braket{ \delta i j k | \hat{O}_{3} | k j i } \ \Big)\\
											&\quad - \sum_{i=1}\sum_{j=1} \epsilon_{ij} \braket{ \delta i | j } \\
											&= \sum_{i} \braket{ \delta i | \hat{O}_{1} | i } + 2 \sum_{i<j} \braket{ \delta i j | \hat{O}_{2} | i j }_{P} \\
											&\quad + 3 \sum_{i<j<k} \braket{ \delta i j k | \hat{O}_{3} | i j k }_{P} - \sum_{i=1}\sum_{j=1} \epsilon_{ij} \braket{ \delta i | j }
\end{split}
\end{equation}
We use the shorthand such that $\ket{ i j .. k }_{P} = \sum_{\hat{P} \in S_{n}} (-1)^{p} \hat{P} \ket{ i j ... k} = \mathcal{P}_{n}\ket{ i j ... k} $, $S_{n}$ being the permutation group of $n$ elements (Equation \ref{TC: Permutations}). Similarly,
\begin{equation}
\begin{split}
	\delta \mathcal{L}_{\phi} 	&= \sum_{i} \braket{ i | \hat{O}_{1} | \delta i } + 2 \sum_{i<j} \braket{ i j | \hat{O}_{2} | \delta i j }_{P} \\
												&\quad + 3 \sum_{i<j<k} \braket{ i j k | \hat{O}_{3} | \delta i j k }_{P} - \sum_{i=1}\sum_{j=1} \epsilon_{ij} \braket{ i | \delta j }
\end{split}
\end{equation}
We seek $\delta \mathcal{L} = 0$ for any arbitrary changes of $\psi$ and $\phi$ independently. Hence, $\delta \mathcal{L}_{\psi} = 0$ and $\delta \mathcal{L}_{\phi} = 0$ independently.

\subsubsection{Effective transcorrelated Hamiltonian}
Consider the condition $\delta \mathcal{L}_{\psi} = 0$.
\begin{equation}
\begin{split}
	\delta \mathcal{L}_{\psi} &= \sum_{i} \braket{ \delta i | \hat{O}_{1} | i } + 2 \sum_{i<j} \braket{ \delta i j | \hat{O}_{2} | i j }_{P} \\
											&\quad + 3 \sum_{i<j<k} \braket{ \delta i j k | \hat{O}_{3} | i j k }_{P} - \sum_{i=1}\sum_{j=1} \epsilon_{ij} \braket{ \delta i | j } \\
											&= \sum_{i} \braket{ \delta i | \hat{h}_{i} | i } + 2 \sum_{i<j} \braket{ \delta i j | ( r_{12}^{-1} -  \hat{K}(\boldsymbol{r}_{1}, \boldsymbol{r}_{2}) ) | i j }_{P} \\
											&\quad +  3 \sum_{i<j<k} \braket{ \delta i j k | \hat{L}(\boldsymbol{r}_{1}, \boldsymbol{r}_{2}, \boldsymbol{r}_{3}) | i j k }_{P} \\
											&\quad - \sum_{i=1}\sum_{j=1} \epsilon_{ij} \braket{ \delta i | j } \\
											&= \sum_{i} \braket{ \delta i | \hat{h}_{i} | i } + \sum_{i=1}\sum_{j=1} \braket{ \delta i j | ( r_{12}^{-1} - \hat{K}(\boldsymbol{r}_{1}, \boldsymbol{r}_{2}) ) | i j }_{P} \\
											&\quad + \frac{1}{2} \sum_{i=1}\sum_{j=1}\sum_{k=1} \braket{ \delta i j k | \hat{L}(\boldsymbol{r}_{1}, \boldsymbol{r}_{2}, \boldsymbol{r}_{3}) | i j k }_{P} \\
											&\quad - \sum_{i=1}\sum_{j=1} \epsilon_{ij} \braket{ \delta i | j } \\
											&= \sum_{i} \Big[ \braket{ \delta i | \hat{h}_{i} | i } + \sum_{j=1} \braket{ \delta i j | ( r_{12}^{-1} -  \hat{K}(\boldsymbol{r}_{1}, \boldsymbol{r}_{2}) ) | i j }_{P} \\
											&\quad +  \frac{1}{2} \sum_{j=1}\sum_{k=1} \braket{ \delta i j k | \hat{L}(\boldsymbol{r}_{1}, \boldsymbol{r}_{2}, \boldsymbol{r}_{3}) | i j k }_{P} \\
											&\quad - \sum_{j=1} \epsilon_{ij} \braket{ \delta i | j } \Big ] \\
											&= 0
\end{split}
\end{equation}
Rewriting the expression more explicitly in integral form,
\begin{equation}
\begin{split}
&\sum_{i} \Big[  \braket { \delta i | \hat{h}_{i} | i } + \sum_{j=1} \braket{ \delta i j | ( r_{12}^{-1} -  \hat{K}(\boldsymbol{r}_{1}, \boldsymbol{r}_{2}) ) | i j }_{P} \\
&+ \frac{1}{2} \sum_{j=1}\sum_{k=1} \braket{ \delta i j k | \hat{L}(\boldsymbol{r}_{1}, \boldsymbol{r}_{2}, \boldsymbol{r}_{3}) | i j k }_{P} \\
&- \sum_{j=1} \epsilon_{ij} \braket{ \delta i | j }  \Big ] = 0
\end{split}
\end{equation}
\begin{equation}
\begin{split}
&\sum_{i=1} \int d\boldsymbol{r}_{1} \delta \psi_{i}^{*}(\boldsymbol{r}_{1}) \Big[ \hspace{0.2em} \hat{h}_{i}(\boldsymbol{r}_{1}) \\
& + \sum_{j=1} \int d\boldsymbol{r}_{2} \psi_{j}^{*}(\boldsymbol{r}_{2} ) ( r_{12}^{-1} -  \hat{K}(\boldsymbol{r}_{1}, \boldsymbol{r}_{2}) ) \mathcal{P}_{2} \phi_{j}(\boldsymbol{r}_{2} ) \\
& + \frac{1}{2}\sum_{j,k=1}\int \int d\boldsymbol{r}_{2} d\boldsymbol{r}_{3} \psi_{j}^{*}(\boldsymbol{r}_{2})\psi_{k}^{*}(\boldsymbol{r}_{3})\hat{L}(\boldsymbol{r}_{1}, \boldsymbol{r}_{2}, \boldsymbol{r}_{3}) \mathcal{P}_{3} \phi_{j}(\boldsymbol{r}_{2})\phi_{k}(\boldsymbol{r}_{3}) \\
& - \sum_{j=1} \epsilon_{ij} \hspace{0.2em} \Big] \phi_{i}(\boldsymbol{r}_{1}) = 0
\end{split}
\end{equation}
Since the expression holds for any $\delta \psi_{i}^{*}$, the terms in the square bracket must be zero. Hence,
\begin{equation}
\begin{split}
&\hat{h}_{i}(\boldsymbol{r}_{1}) + \sum_{j=1} \int d\boldsymbol{r}_{2} \psi_{j}^{*}(\boldsymbol{r}_{2} ) ( r_{12}^{-1} -  \hat{K}(\boldsymbol{r}_{1}, \boldsymbol{r}_{2}) ) \mathcal{P}_{2} \phi_{j}(\boldsymbol{r}_{2} ) \\
& + \frac{1}{2}\sum_{j,k=1} \int \int d\boldsymbol{r}_{2} d\boldsymbol{r}_{3} \psi_{j}^{*}(\boldsymbol{r}_{2})\psi_{k}^{*}(\boldsymbol{r}_{3})\hat{L}(\boldsymbol{r}_{1}, \boldsymbol{r}_{2}, \boldsymbol{r}_{3}) \mathcal{P}_{3} \phi_{j}(\boldsymbol{r}_{2})\phi_{k}(\boldsymbol{r}_{3}) \\
& - \sum_{j=1} \epsilon_{ij} = 0
\end{split}
\end{equation}
for all $i$.
Rearrangement of the equation recovers the equation given in equation \ref{TC: Effective Hamiltonian}.

\section{Correlator parameters}
\label{Appendix B}

\subsection{Correlator parameters for closed shell atoms (SOM8)}

The correlator parameters for closed shell ions (Table \ref{SOM-CS}) found from using SOM minimisation on a set of 8 parameters are tabulated in Table \ref{SOM-CS-Appendix}.
\begin{table*}[h!]
\centering
\begin{tabular}{c c c c c c c c }
\hline
\hline
		$m$		&			$n$		&			$o$		&					He				&					Be					&					C					&					O					&					Ne				\\
\hline
		0			&			0			&			1			&				0.50000			&				0.50000			&				0.50000			&				0.50000			&				0.50000			\\
		0			&			0			&			2			&				0.42476			&				0.21162			&				0.03755			&			  -0.45608			&			  -0.75272			\\
		0			&			0			&			3			&			  -0.23302			&			    0.30937			&			  -0.18384			&			    0.67271			&			   1.31436			\\
		0			&			0			&			4			&				0.28445			&			  -0.21062			&				0.42119			&				0.04159 			&			  -0.36159 			\\
		1			&			0			&			0			&				0.00151 			&			    0.01082			&			  -0.00952			&			  -0.02441			&			  -0.00979			\\
		2			&			0			&			0			&			  -0.16865			&			  -0.11872			&			  -0.12690 			&			  -0.12667			&			  -0.14499			\\
		3			&			0			&			0			&			  -0.34421			&			  -0.17257			&			  -0.05827			&			  -0.01992			&			  -0.00973			\\
		4			&			0			&			0			&			  -0.54727			&				0.16579			&				0.08346			&				0.02964			&				0.08552			\\
\hline				
\hline
\end{tabular}
\caption{Correlator parameters for closed shell atoms found from using SOM minimisation. These parameters are used to obtain the transcorrelated energies in Table \ref{SOM-CS}.}
\label{SOM-CS-Appendix}
\end{table*}

\subsection{Correlator parameters for helium from different initial guesses (SOM8)}
The correlator parameters for helium (Table \ref{Results: Different guesses}) found from using different initial guesses are tabulated in Table \ref{Different guesses-Appendix}.
\begin{table*}
\centering
\begin{tabular}{c c c c c c c c }
\hline
\hline
		$m$		&			$n$		&			$o$		&		$c_{100} = +1$		&		$c_{100} = -1$		&			$c_{100} = -2$	\\
\hline
		0			&			0			&			1			&				0.50000			&				0.50000			&				0.50000			\\
		0			&			0			&			2			&				0.29627			&				0.29152			&				0.29071			\\
		0			&			0			&			3			&			    0.12627			&			    0.14407			&			    0.15048			\\
		0			&			0			&			4			&				0.04485			&			    0.06602			&				0.07330			\\
		1			&			0			&			0			&				0.99565 			&			  -1.00073			&			  -1.99929			\\
		2			&			0			&			0			&			  -0.00338			&			  -0.00177			&			  -0.00116 			\\
		3			&			0			&			0			&			  -0.00200			&			  -0.00159			&			  -0.00134			\\
		4			&			0			&			0			&			  -0.00110			&			  -0.00101			&			  -0.00100			\\
\hline				
\hline
\end{tabular}
\caption{Correlator parameters for helium found from using SOM minimisation with different initial guesses. These parameters are used to obtain the transcorrelated energies in table \ref{Results: Different guesses}.}
\label{Different guesses-Appendix}
\end{table*}

\subsection{Correlator parameters for helium-like ions (SOM18)}
The correlator parameters for helium-like ions found from using SOM minimisation on a set of 18 parameters are tabulated in Tables \ref{HF-SOM-Appendix1} and \ref{HF-SOM-Appendix2}. The parameters for helium were used as a starting guess for the helium-like systems in Table \ref{Results: HF-SOM}.
\begin{table*}
\centering
\begin{tabular}{c c c c c c c c }
\hline
\hline
	$m$			&			$n$		&			$o$		&				H$^{-}$			&					He				&				Li$^{+}$			&				Be$^{2+}$		&				B$^{3+}$			\\
\hline
		0			&			0			&			1			&				0.50000			&				0.50000			&				0.50000			&				0.50000			&				0.50000			\\
		0			&			0			&			2			&				0.17646			&				0.10188			&				0.04101			&			  -0.02237			&			  -0.09254			\\
		0			&			0			&			3			&			  -0.33069			&			  -0.38197			&			  -0.40395			&			  -0.42278			&			  -0.44691			\\
		0			&			0			&			4			&				0.99262			&				0.95942			&				0.95251			&				0.94670			&				0.93836			\\
		1			&			0			&			0			&				0.00594			&			  -0.00015			&				0.00225			&				0.00645			&				0.01269			\\
		2			&			0			&			0			&				0.23633			&				0.23188			&				0.23339			&				0.23499			&				0.23692			\\
		3			&			0			&			0			&			  -0.44818			&			  -0.45048			&			  -0.44978			&			  -0.44932			&			  -0.44888			\\
		4			&			0			&			0			&				0.82844			&				0.82766			&				0.82795			&				0.82804			&				0.82811			\\
		2			&			2			&			0			&			  -4.16199			&			  -4.15465			&			  -4.15276			&			  -4.15252			&			  -4.15303			\\
		2			&			0			&			2			&				0.84241			&				0.80798			&				0.80157 			&				0.79821 			&				0.79362			\\
		2			&			2			&			2			&			  10.19802			&			  10.19694			&			  10.19694			&			  10.19683			&			  10.19663			\\
		4			&			0			&			2			&			  -4.95213			&			  -4.96235			&			  -4.96281			&			  -4.96305			&			  -4.96336			\\
		2			&			0			&			4			&			  -1.34054			&			  -1.35616			&			  -1.35651			&			  -1.35712			&			  -1.35804			\\
		4			&			2			&			2			&			  -5.91087			&			  -5.90919			&			  -5.90919			&			  -5.90931			&			  -5.90940			\\
		6			&			0			&			2			&				0.90672			&			    0.90347			&				0.90345			&				0.90341			&				0.90338			\\
		4			&			0			&			4			&				5.51230			&			 	5.50742			&				5.50747			&				5.50736			&				5.50726			\\
		2			&			2			&			4			&			  -0.03148			&			  -0.03160			&			  -0.03156			&			  -0.03163			&			  -0.03170			\\
		2			&			0			&			6			&			  -1.04601			&			  -1.05188			&			  -1.05165			&			  -1.05186			&			  -1.05208			\\
		
\hline				
\hline
\end{tabular}
\caption{Correlator parameters for helium-like ions (H$^{-}$ to B$^{3+}$) found from using SOM minimisation.}
\label{HF-SOM-Appendix1}
\end{table*}

\begin{table*}
\centering
\begin{tabular}{c c c c c c c c }
\hline
\hline
	$m$			&			$n$		&			$o$	&				C$^{4+}$			&				N$^{5+}$			&				O$^{6+}$			&				F$^{7+}$			&				Ne$^{8+}$		\\
\hline
		0			&			0			&			1			&				0.50000			&				0.50000			&				0.50000			&				0.50000			&				0.50000			\\
		0			&			0			&			2			&			  -0.16520 			&			  -0.24505 			&			  -0.21726			&			  -0.28297			&			  -0.40323			\\
		0			&			0			&			3			&			  -0.47290			&			    0.92623 			&			  -2.85437 			&			  -2.72267 			&			  -2.81519			\\
		0			&			0			&			4			&				0.92969			&				0.02574			&			  -0.22549			&			  -0.03555			&			  -0.07534 			\\
		1			&			0			&			0			&				0.02024			&			  	0.23972			&				0.73292 			&				0.78108			&				0.63683  			\\
		2			&			0			&			0			&				0.23880			&			  -0.44851			&				0.42552			&				0.41999 			&				0.37976			\\
		3			&			0			&			0			&			  -0.44855			&			    0.82809			&			  -0.40369			&			  -0.40880			&			  -0.41889			\\
		4			&			0			&			0			&				0.82814			&			  -4.15395			&				0.83915			&				0.83706			&				0.83451			\\
		2			&			2			&			0			&			  -4.15379			&			  -4.15465			&			  -4.27033			&			  -4.24468			&			  -4.24557			\\
		2			&			0			&			2			&				0.78910			&				0.78741			&				0.17596			&				0.29033			&				0.25548			\\
		2			&			2			&			2			&			  10.19648			&			  10.19646    		&			  10.18118			&			  10.18753			&			  10.18630			\\
		4			&			0			&			2			&			  -4.96357			&			  -4.96347			&			  -5.00535 			&			  -4.99058 			&			  -4.99174			\\
		2			&			0			&			4			&			  -1.35874			&			  -1.35859			&			  -1.47410			&			  -1.43261			&			  -1.43795			\\
		4			&			2			&			2			&			  -5.90944			&			  -5.90945			&			  -5.91112			&			  -5.91019			&			  -5.91031			\\
		6			&			0			&			2			&				0.90337			&			    0.90340			&				0.89998			&				0.90163			&				0.90157			\\
		4			&			0			&			4			&				5.50723			&			 	5.50727			&				5.49931			&				5.50322			&				5.50289			\\
		2			&			2			&			4			&			  -0.03174			&			  -0.03174			&			  -0.03402			&			  -0.03279			&			  -0.03297			\\
		2			&			0			&			6			&			  -1.05219			&			  -1.05213			&			  -1.06807			&			  -1.06012			&			  -1.06121			\\
		
\hline				
\hline
\end{tabular}
\caption{Correlator parameters for helium-like ions (C$^{4+}$ to Ne$^{8+}$) found from using SOM minimisation.}
\label{HF-SOM-Appendix2}
\end{table*}

\end{multicols}
\clearpage
\begin{multicols}{2}
\printbibliography
\end{multicols}
\end{document}